\documentclass[11pt,a4paper]{article}

\pdfoutput=1

\usepackage{jheppub}
\usepackage[utf8]{inputenc}
\usepackage[dvipsnames]{xcolor}
\usepackage[T1]{fontenc}
\usepackage{tocloft}
\usepackage[normalem]{ulem}
\usepackage{subcaption}
\usepackage{array}
\usepackage{enumitem}
\usepackage{extarrows}
\usepackage{mathtools}
\usepackage{lipsum}
\usepackage{tikz}
\usepackage{tikz-3dplot}
\usepackage{pgfplots}

\pgfplotsset{compat=newest}

\usepgfplotslibrary{external} 
\usetikzlibrary{decorations.pathmorphing}
\usetikzlibrary{decorations.markings}
\usetikzlibrary{decorations.pathreplacing}
\usetikzlibrary{arrows,automata,calc,shapes,matrix,trees,positioning,through,chains}
\usetikzlibrary{fadings}

\allowdisplaybreaks[1] 

\newcommand{\softserve}{{\tt SoftSERVE}}
\newcommand{\softserveonine}{{\tt SoftSERVE\;0.9}}
\newcommand{\softserveone}{{\tt SoftSERVE\;1.0}}

\newcommand{\eps}{\epsilon}
\newcommand{\taubar}{\bar\tau}
\newcommand{\no}{\nonumber}
\newcommand{\altsqrt}[1]{\sqrt{\rule{0mm}{3.5mm} #1}}
\def\rd{\mathrm{d}}

\title{\boldmath 
Generic dijet soft functions at two-loop order:
uncorrelated emissions}

\author[a]{Guido Bell,}
\author[b,c,d]{Rudi Rahn,}
\author[e,f]{and Jim Talbert}

\note{SI-HEP-2020-11, QFET-2020-01, Nikhef 2020-007, DESY 19-157.}

\affiliation[a]{Theoretische Physik 1, Naturwissenschaftlich-Technische Fakult\"at, 
Universit\"at Siegen,\\Walter-Flex-Stra{\ss}e 3, 57068 Siegen, Germany}
\affiliation[b]{Institute for Theoretical Physics Amsterdam and Delta Institute for Theoretical Physics,\\
University of Amsterdam, Science Park 904, 1098 XH, Amsterdam, The Netherlands}
\affiliation[c]{Nikhef, Theory Group, Science Park 105, 1098 XG, Amsterdam, The Netherlands}
\affiliation[d]{Albert Einstein Center for Fundamental Physics, 
Institut f\"ur Theoretische Physik,\\
Universit\"at Bern, Sidlerstrasse 5, 3012 Bern, Switzerland}
\affiliation[e]{Niels Bohr International Academy, Niels Bohr Institute, University of Copenhagen,\\
Blegdamsvej 17, DK-2100 Copenhagen, Denmark}
\affiliation[f]{Theory Group, Deutsches Elektronen-Synchrotron (DESY), Notkestra{\ss}e 85,\\
22607 Hamburg, Germany}

\emailAdd{bell@physik.uni-siegen.de}
\emailAdd{rudi.rahn@uva.nl}
\emailAdd{ronald.talbert@nbi.ku.dk}

\abstract{We extend our algorithm for automating the calculation of two-loop dijet soft functions 
to observables that do not obey the non-Abelian exponentiation theorem, i.e. to those that require 
an independent calculation of the uncorrelated-emission contribution. As the singularity structure
of uncorrelated double emissions differs substantially from the one for correlated emissions, we 
introduce a novel phase-space parametrisation that isolates the corresponding divergences. The 
resulting integrals are implemented in \softserveone, which we release alongside of this work, 
and which we supplement by a regulator that is consistent with the rapidity renormalisation 
group framework. Using our automated setup, we confirm existing results for various jet-veto 
observables and provide a novel prediction for the soft-drop jet-grooming algorithm.}

\keywords{QCD, Soft-Collinear Effective Theory (SCET), NNLO Computations}

\begin{document} 

\maketitle
\flushbottom

\section{Introduction}
\label{sec:intro}

The perturbative calculation of soft functions provides insights into the infrared structure of
gauge theory amplitudes and enables the resummation of logarithmically enhanced corrections
to all orders in perturbation theory. Starting at next-to-next-to-leading order (NNLO) and
beyond, the perturbative computations often become intricate since the divergences in the 
phase-space integrations overlap. This motivated us to develop a systematic algorithm for 
the calculation of two-loop soft functions in~\cite{Bell:2018vaa,Bell:2018oqa}, which exploits 
the fact that the defining matrix element of the soft functions is independent of the observable
for a given hard-scattering process.

In this work we are concerned with soft functions that arise in processes with two massless, 
coloured, hard partons that are in a back-to-back configuration. These \emph{dijet soft functions}
can be defined in terms of two light-like Wilson lines $S_{n}$ and $S_{\bar{n}}$, which embed 
the eikonal form of the soft interactions and which trace the directions $n^{\mu}$ and 
$\bar{n}^{\mu}$ of the (initial or final-state) hard partons with $n^2=\bar{n}^2=0$ and 
$n\cdot{\bar{n}}=2$. A generic soft function of this type can be written in
the form
\begin{align}
S(\tau, \mu) = \frac{1}{N_c} \; \sum_{i\in X} \;
\mathcal{M}(\tau;\lbrace k_{i} \rbrace)\;
\mathrm{Tr}\; 
|\langle X | \,T [S^{\dagger}_{n}(0) S_{\bar{n}}(0)]\, | 0 \rangle |^{2}\,,
\label{eq:softfun:definition}
\end{align}
where $\mathcal{M} \left( \tau ; \lbrace k_{i} \rbrace \right)$ represents an observable-specific 
measurement on the soft radiation $X$ with partonic momenta $\lbrace k_{i} \rbrace$, which -- after
isolating the singularities present in the soft matrix element -- acts as a weight factor for the 
phase-space integrations. 

In~\cite{Bell:2018oqa} we specified a number of constraints that we impose on the functional form 
of the measure\-ment function $\mathcal{M} \left( \tau ; \lbrace k_{i} \rbrace \right)$, 
whose resulting generality was illustrated in applications to about a dozen $e^+e^-$ and 
hadron-collider soft functions. What all of these observables have in common is that they are 
consistent with non-Abelian exponentiation (NAE)~\cite{Gatheral:1983cz,Frenkel:1984pz}. 
In a non-Abelian gauge theory this implies that for any observable the all-order soft matrix 
element takes the form of an exponential, which involves only Feynman diagrams with specific 
colour structures. At NNLO this fixes one of the three colour structures to the square of the 
NLO amplitude, and as long as the measure\-ment function itself factorises into two 
single-emission pieces, this contribution to the soft function does not require a dedicated 
calculation since it is proportional to the square of the NLO soft function. This allowed us 
in~\cite{Bell:2018oqa} to present complete results for NAE observables, although the algorithm 
devised in that paper applies only to two out of three NNLO colour structures, which constitute 
the so-called correlated-emission contribution.

There exist, however, interesting soft functions that do not comply with NAE, and which require 
an independent calculation of the uncorrelated-emission contribution. This applies, for instance, 
to soft functions that are defined in terms of a jet algorithm, which partitions the phase space 
of the soft emissions into different regions in which the partons are clustered together. 
As these clustering constraints do not have an analogue at lower orders, the respective measurement 
function does not factorise into single-emission pieces and the uncorrelated-emission 
contribution becomes non-trivial. 

The singularity structure of uncorrelated double emissions differs, on the other hand, from 
the one for correlated emissions, and the phase-space parametrisation we used 
in~\cite{Bell:2018oqa} fails to factorise the corresponding divergences. At first sight one may 
think that the calculation of the uncorrelated-emission contribution should be simpler than the one 
for correlated emissions since the underlying matrix element is trivial. As we will see in this 
paper, however, the singularity structure imposes more stringent constraints on the required 
phase-space parametrisation in a generic, observable-independent approach. It therefore turns 
out that one cannot apply a universal parametrisation for all observables in this case, 
but one instead has to resort to specific parametrisations for different classes of soft functions. 
We actually already presented the phase-space parametrisation we use for uncorrelated emissions 
in~\cite{Bell:2018vaa,Bell:2018jvf}, in which we focused on the divergences of the soft functions, 
whereas we present complete NNLO results in this paper.

Apart from devising a systematic algorithm for the calculation of dijet soft functions, we 
developed a stand-alone program called \softserve~for their numerical 
evaluation~\cite{Bell:2018oqa}. Whereas the previous version \softserveonine~could only be used 
for the calculation of the correlated-emission contribution, the new version \softserveone~-- 
which we publish alongside this paper -- contains a number of new features. Most importantly, 
we implemented the master formula derived in this work for the calculation of the 
uncorrelated-emission contribution, such that \softserveone~can now handle generic dijet soft 
functions that comply with our ansatz, both for NAE and NAE-violating observables. Moreover, the
new version contains a script for the renormalisation of cumulant soft functions, which differs 
from the one for Laplace-space soft functions considered in~\cite{Bell:2018oqa}, and we implemented 
the formulae from~\cite{Bell:2018vaa}, which allow for a direct calculation of the soft anomalous 
dimension (and also the collinear anomaly exponent~\cite{Becher:2010tm,Becher:2011pf}), without 
having to calculate the complete bare soft function. Finally, we argued in~\cite{Bell:2018oqa} 
that the rapidity regulator that is used in \softserveonine~is not suited for the rapidity 
renormalisation group (RRG) approach~\cite{Chiu:2012ir}, since it is not implemented on the level of 
connected webs. In the new version we remedied this point by adding an option which allows the 
user to run \softserve~with different rapidity regulators. Whereas we briefly comment on all of 
these changes in this work, we refer to the \softserve~user manual for more detailed explanations. 
The \softserve~distribution is publicly available at \url{https://softserve.hepforge.org/}.

The remainder of the paper develops as follows: in Section~\ref{sec:measure} we introduce the 
phase-space parametrisation we use for uncorrelated emissions as well as the
cor\-responding form of the measurement function. In Section~\ref{sec:baresoftfun} we employ 
this parametrisation to obtain a master formula for the calculation of the uncorrelated-emission 
contribution to a generic bare two-loop soft function, which we then renormalise in 
Section~\ref{sec:renormalize}. In Section~\ref{sec:softserve} we briefly review the technical 
aspects of the \softserve~extension, and we present sample results for NAE and NAE-violating 
observables in Section~\ref{sec:results}, including a novel calculation of an NNLO soft function 
for the soft-drop jet-grooming algorithm.  We finally conclude in Section~\ref{sec:conclude}, and 
we present some technical aspects of our analysis in an appendix.

\section{Measurement function}
\label{sec:measure}

Following the procedure outlined in~\cite{Bell:2018oqa}, we restrict ourselves to soft functions 
whose defining measurements are of the form
\begin{align}
\mathcal{M}(\tau;\lbrace k_{i} \rbrace) = 
\exp\big(-\tau\, \omega(\lbrace k_{i} \rbrace)\,\big)\,,
\label{eq:measure:general}
\end{align}
where it is clear from the exponential that we typically evaluate the soft functions in some space 
conjugate to momentum space, e.g.~Laplace or Fourier space. The variable $\tau$ then denotes the 
associated conjugate variable, and the function $\omega(\lbrace k_{i} \rbrace)$ characterises 
the specific constraint on the final-state momenta that is provided by the observable in
question.  More specifically, we assume that
\begin{enumerate}[label=\bf(A\arabic*)]
\item
the soft function is embedded in a dijet factorisation theorem and it has a double-logarithmic 
evolution in the renormalisation scale $\mu$ and, possibly, also the rapidity scale $\nu$;
\item
$\Re\big(\omega(\lbrace k_{i} \rbrace)\big)\geq0$ and $\omega(\lbrace k_{i} \rbrace)$ is  allowed 
to vanish only for configurations with zero weight in the phase-space integrations, and it is 
furthermore supposed to be independent of the dimensional and the rapidity regulators;
\item
the variable $\tau$ has dimension 1/mass;
\item
the function $\omega(\lbrace k_{i} \rbrace)$ is symmetric under 
$n^\mu \leftrightarrow \bar{n}^{\mu}$ exchange; 
\item
the soft function depends only on one variable $\tau$ in conjugate space, although we already 
showed in~\cite{Bell:2018oqa} how to relax this condition, which is needed e.g.~for 
multi-differential soft functions;
\item
the function $\omega(\lbrace k_{i} \rbrace)$ depends only on one angle $\theta_i$ per emission 
in the $(d-2)$-dimen\-sional transverse plane to $n^\mu$ and $\bar{n}^{\mu}$ as well as on 
relative angles $\theta_{ij}$ between two emissions.
\end{enumerate}
For further explanations regarding these assumptions, we refer the reader to the corresponding 
section in~\cite{Bell:2018oqa}.

In order to illustrate the implications of these assumptions, we considered three template 
observables in~\cite{Bell:2018oqa} relevant for $e^+e^-$ event shapes, threshold resummation and 
trans\-ver\-se-momentum resummation that are all consistent with NAE. We find it convenient to 
proceed similarly in this work, and to highlight the salient features of NAE-violating observables
with a specific example. To this end, we consider the C-parameter-like jet veto observable 
$\mathcal{T}_{C\rm cm}$ from~\cite{Gangal:2014qda}, whose measurement function in Laplace space 
can be written in the form \eqref{eq:measure:general}, except for a global factor of $1/\tau$
which arises because the constraint on the soft radiation is given in momentum space in the form of 
a $\theta$-function rather than a $\delta$-function. This factor is typical for cumulant soft 
functions, and we will investigate its consequences more closely when we discuss renormalisation 
in Section~\ref{sec:renormalize}. For the calculation of the bare soft function, however, this 
factor is just a constant and can be ignored.

For zero and one emissions, the observable is just the usual C-parameter event shape, which we
discussed at length in~\cite{Bell:2018oqa}. The clustering constraint, on the other hand, only
becomes relevant for two and more emissions. Specifically for two emissions with momenta $k$ and
$l$, we have 
\begin{align}
\omega^{CPV}(R;k,l) &= \theta(\Delta-R)\; \max\bigg(\frac{k_+ k_-}{k_++k_-},
\frac{l_+ l_-}{l_++l_-}\bigg) + \theta(R-\Delta )\; 
\frac{(k_++l_+)(k_-+l_-)}{k_++l_++k_-+l_-}\,,
\label{eq:omega:CPveto}
\end{align}
where we introduced light-cone coordinates via $k_+ = n \cdot k$ and $k_- = \bar n \cdot k$, and
\begin{align}
\Delta= \sqrt{\frac14 \ln^2 \frac{k_-l_+}{k_+l_-}+\theta_{kl}^2}
\end{align}
represents the distance measure of the jet algorithm. From \eqref{eq:omega:CPveto} we see
that emissions that are closer than the jet radius $R$ are clustered together, whereas those 
that are further apart are treated as individual emissions, such that the jet veto constrains 
the one with a larger value of the C-parameter. One easily verifies that the assumptions 
(A1)-(A6) are satisfied for this observable, and from \eqref{eq:omega:CPveto} it is obvious 
that $\omega^{CPV}(R;k,l)$ cannot be written as a sum of single-emission functions, which
would be required for a factorisation of the measurement function \eqref{eq:measure:general}.
The observable therefore violates NAE.

In analogy to the correlated-emission calculation from~\cite{Bell:2018oqa}, we need to find a 
parametrisation of the double-emission measurement function that has a well-defined behaviour 
in the singular limits of the corresponding matrix element. The parametrisation we use for 
uncorrelated emissions was already given in~\cite{Bell:2018jvf},
\begin{align}
y_k &= \frac{k_{+}}{k_{-}}\,, \hspace{1.2cm}
q_T = \sqrt{k_+ k_-} \left( \frac{l_-+l_+}{\sqrt{l_+ l_-}} \right)^{n}
+ \sqrt{l_+ l_-} \left( \frac{k_-+k_+}{\sqrt{k_+ k_-}} \right)^{n},
\nonumber\\[0.2em]
y_l &=\frac{l_{+}}{l_{-}} \,, \hspace{1.5cm}
b = \left(\frac{k_{+}k_{-}}{l_{+}l_{-}}\right)^{\frac{n+1}{2}} 
\left( \frac{l_-+l_+}{k_-+k_+} \right)^{n},
\label{eq:parametr:nnlo}
\end{align}
where $n$ is a parameter that is related to the power counting of the modes in the effective 
theory -- see the discussion in~\cite{Bell:2018oqa}. Unlike the correlated-emission case, 
we thus use specific parametrisations for classes of observables that correspond to the same 
value of $n$. The parametrisation becomes, for instance, particularly simple for SCET-2 soft 
functions where $n=0$.

In physical terms, the variables $y_k$ and $y_l$ are measures of the rapidities of the 
individual partons, whereas $b$ and $q_T$ only have a simple interpretation for $n=0$, 
where they correspond to the ratio and the scalar sum of their transverse momenta, 
respectively (the $n$-dependent terms introduce rapidity-dependent weight factors). 
Similar to~\cite{Bell:2018oqa}, the parametrisation is supplemented by the angular variables
\begin{align}
t_k = \frac{1-\cos\theta_k}{2}\,, \qquad\qquad 
t_l = \frac{1-\cos\theta_l}{2}\,, \qquad\qquad
t_{kl} = \frac{1-\cos\theta_{kl}}{2}\,,
\label{eq:parametr:nnlo:angles}
\end{align}
with $\theta_k=\sphericalangle(\vec{v}_\perp,\vec{k}_\perp)$, 
$\theta_l=\sphericalangle(\vec{v}_\perp,\vec{l}_\perp)$ and
$\theta_{kl}=\sphericalangle(\vec{k}_\perp,\vec{l}_\perp)$. The vector $v^\mu$ encodes a 
potential azimuthal dependence of the observable around the collinear axis -- see~\cite{Bell:2018oqa} 
for specific examples. The inverse trans\-formation to \eqref{eq:parametr:nnlo} can be found  
in~\cite{Bell:2018vaa}.

The integration ranges for the variables $y_k$, $y_l$ and $b$ span the entire positive real 
axis and, similar to the correlated-emission case, they can be mapped onto the unit hypercube
using symmetry arguments. The implicit phase-space divergences then arise in the 
following four limits:
\begin{itemize}
\item
$q_T\rightarrow 0$, which corresponds to the situation in which both emitted partons become soft;
\item
$b \rightarrow 0$, which implies that the parton with momentum $k^\mu$ becomes soft
(compared to $l^\mu$);
\item
$y_k \rightarrow 0$, which reflects the fact that the parton with momentum $k^\mu$ becomes 
collinear to the direction $n^\mu$ (at fixed transverse momentum);
\item
$y_l \rightarrow 0$, which is the corresponding limit for the parton with momentum $l^\mu$.
\end{itemize}
As $q_T$ is the only dimensionful variable in our parametrisation and the mass dimen\-sion 
of the variable $\tau$ is fixed by (A3), the function 
$\omega(\lbrace k,l\rbrace)=\omega(q_T,y_k,y_l,b,t_k,t_l,t_{kl})$ 
must be linear in $q_T$. The limit $b\to 0$ is furthermore protected by infrared safety, 
which means that the measurement function cannot vanish in this limit since it must fall back 
to the one-emission function~\cite{Bell:2018oqa}. Yet, we still have to control the measurement 
function in the remaining two limits to make sure that we can properly extract the associated 
divergences. 

The very fact that one has to control the measurement function in two unprotected singular 
limits -- as opposed to one for correlated emissions -- is the main complication in the
present calculation. To better illustrate this point, let us for the moment consider a
generic observable that obeys NAE, i.e.~its two-emission measurement function can be written
in the form
\begin{align}
\mathcal{M}_2(\tau; k,l) &=\mathcal{M}_1(\tau; k) \; \mathcal{M}_1(\tau;l) \,,
\nonumber\\[0.2em]
&=  \exp\Big\{-\tau\, \Big( k_{T}\, y_k^{n/2}\, f(y_k,t_k)
+ l_{T}\, y_l^{n/2}\, f(y_l,t_l) \Big)\,\Big\},
\label{eq:measure:NAE}
\end{align}
where we have used the explicit form of the single-emission measurement function 
from~\cite{Bell:2018oqa}, and the function $f(y,t)$ is by construction finite and non-zero
as $y\to 0$. In order to extract the collinear divergences that arise in the limits 
$y_k \rightarrow 0$ and $y_l \rightarrow 0$, one has to make sure that the term in the round 
parenthesis is finite and non-zero in either of the limits and in the combined limit 
$y_k,y_l\to 0$ as well. Except for $n=0$ this is obviously not the case. Factoring out 
$y_k^{n/2}$, on the other hand, would guarantee that the first term stays finite as $y_k\to 0$, 
but at the same time the second term would blow up for $n>0$. Similarly, factoring out powers 
of $y_l$ does not help to make the expression in the parenthesis finite as $y_l\to 0$.

The problem is solved by the specific form of the parametrisation \eqref{eq:parametr:nnlo}. 
In terms of these variables, the transverse-momentum variables $k_T$ and $l_T$ take the form
\begin{align}
k_T = \sqrt{k_+ k_-} = \left( \frac{\sqrt{y_l}}{1+y_l} \right)^{n}
\frac{b}{1+b} \;\,q_{T}\,, 
\qquad
l_T = \sqrt{l_+ l_-} = \left( \frac{\sqrt{y_k}}{1+y_k} \right)^{n}
\frac{1}{1+b} \;\,q_{T}\,,
\end{align}
which -- when inserted into \eqref{eq:measure:NAE} -- shows that both terms in the parenthesis
are proportional to $y_k^{n/2}y_l^{n/2}$. Once this term is factored out, the remaining 
expression is thus finite and non-zero in the collinear limits as desired. This explains why 
the phase-space parametrisation for uncorrelated emissions must be $n$-dependent, and it motivates 
the following ansatz for the double-emission measurement function:
\begin{align}
\label{eq:measure:NNLO:unc}
\mathcal{M}_2^{unc}(\tau; k,l) = \exp\big(-\tau\, q_{T}\, y_k^{n/2}\, y_l^{n/2}\, 
G(y_k,y_l,b,t_k,t_l,t_{kl})\,\big)\,,
\end{align}
where the dependence on $q_T$ is fixed on dimensional grounds and the function $G$ is supposed 
to be finite and non-zero as $y_k\to0$ and $y_l\to0$. Although our discussion started from the 
specific form \eqref{eq:measure:NAE} of a NAE observable, we expect that generic NAE-violating 
observables can be written in the form \eqref{eq:measure:NNLO:unc} as well. The reason is that 
the soft function is by assumption embedded in a dijet factorisation theorem -- see (A1) --
and the pole cancellation between the various regions requires that a potential NAE-violating 
term in the two-emission measurement function cannot upset the scaling in the limits
$y_k\to0$ and $y_l\to0$. The discussion is actually similar to the one in Appendix~A
of~\cite{Bell:2018oqa}. 

As an example we consider the jet-veto template from above, which corresponds
to $n=1$, $f(y_k,t_k)= 1/(1+y_k)$ and
\begin{align}
\label{eq:G:CPveto}
 G(y_k,y_l,b,t_k,t_l,t_{kl}) &=
 \theta(\Delta_G-R)\; \frac{\max(1,b)}{(1+b)\,(1+y_k)\,(1+y_l)}
\\[0.3em]
&\quad + \theta(R-\Delta_G )\; 
\frac{\big(1+y_l+(1+y_k)b\big)\,\big(y_k(1+y_l)+b y_l(1+y_k)\big)}
{\big(1+b\big)\,\big(1+y_k\big)\,\big(1+y_l\big)\,\big(y_k(1+y_l)^2+b y_l(1+y_k)^2\big)}\,,
\nonumber
\end{align}
where the distance measure is now given by
\begin{align}
\label{eq:Delta}
\Delta_G= \sqrt{\frac14 \ln^2 \frac{y_k}{y_l}+\arccos^2(1-2t_{kl})}\,.
\end{align}
Due to the factorisation of $\sqrt{y_k y_l}$ in \eqref{eq:measure:NNLO:unc}, we see that the 
expression in \eqref{eq:G:CPveto} is indeed finite in the limits $y_k\to0$ and $y_l\to0$ as
required. The distance measure \eqref{eq:Delta} reveals, moreover, that the precise form in 
which the collinear limits are evaluated matters, and we will come back to this point at the end 
of this section.

Before doing so, we analyse the general constraints on the double-emission measure\-ment 
function that arise from infrared safety. Following~\cite{Bell:2018oqa}, we express the 
variables $b$ and $q_T$ in terms of those that parametrise the one-particle phase space
for each of the emitted partons,
\begin{align}
b = \frac{k_T}{l_T} \left( \frac{\sqrt{y_k}}{1+y_k} \, 
\frac{1+y_l}{\sqrt{y_l}} \right)^{n}, \qquad
q_T = k_T \left( \frac{1+y_l}{\sqrt{y_l}} \right)^{n} 
+ l_T \left( \frac{1+y_k}{\sqrt{y_k}} \right)^{n}.
\end{align}
The limit in which the parton with momentum $k^\mu$ becomes soft then corresponds to $k_T\to 0$,
which translates into $b\to 0$ and $q_T\to l_T \big((1+y_k)/\sqrt{y_k}\big)^n$. Infrared safety 
implies that the double-emission measurement function is related to the single-emission 
function in this limit, which yields
\begin{align}
G(y_k,y_l,0,t_k,t_l,t_{kl}) = \frac{f(y_l,t_l)}{(1+y_k)^n}\,.
\label{eq:infrared:soft}
\end{align}
As stated above, this relation guarantees that the function $G$ does not vanish in one of the 
singular limits of the uncorrelated-emission contribution. One can derive a similar constraint 
in the limit in which the two emitted partons become collinear to each other, and in this case 
one finds
\begin{align}
G(y_l,y_l,b,t_l,t_l,0) = \frac{f(y_l,t_l)}{(1+y_l)^n}\,.
\label{eq:infrared:collinear}
\end{align}
Relations \eqref{eq:infrared:soft} and \eqref{eq:infrared:collinear} reflect the fact that the 
observable is infrared safe, and they can easily be checked explicitly for the 
jet-veto template from above.

As already mentioned, we find it convenient to map the integration region onto the unit hypercube 
using symmetry arguments under $n\leftrightarrow \bar n$ and $k\leftrightarrow l$ exchange. Similar 
to~\cite{Bell:2018oqa}, this comes at the price of introducing two different versions of the 
measurement function, which we label by the letters ``A'' and ``B''. As we will explain in more 
detail in Section~\ref{sec:baresoftfun}, they are given by
\begin{align}
G_A(y_k,y_l,b,t_k,t_l,t_{kl}) &= G(y_k,y_l,b,t_k,t_l,t_{kl})\,,
\nonumber\\[0.2em]
G_B(y_k,y_l,b,t_k,t_l,t_{kl}) &= \begin{dcases}          
\,y_k^{-n} \;G(1/y_k,y_l,b,t_k,t_l,t_{kl}) &\,\text{or} \\
\,y_l^{-n} \;G(y_k,1/y_l,b,t_k,t_l,t_{kl})\,.
\end{dcases}
\end{align}
Physically, region A corresponds to the case in which both partons are emitted into the same 
hemisphere with respect to the collinear axis, whereas region B describes the opposite-hemisphere 
case.

Finally, we saw in \eqref{eq:Delta} that the distance measure of the jet algorithm is ambiguous 
in the double limit $y_k\to0$ and $y_l\to0$, since it matters if the limit is evaluated at a fixed 
ratio $y_k/y_l$ or if it is evaluated sequentially. Physically, this corresponds to a distinction 
between the joint collinear limit of the emitted partons at a fixed rapidity distance and the 
individual collinear limits of each of the partons. The ambiguity only arises in the same-hemisphere 
case, and it can be disentangled via a sector decomposition strategy. As we will show in the next 
section, this introduces two subregions in region A with
\begin{align}
G_{A_1}(y,r,b,t_k,t_l,t_{kl}) &= G_A(y,r y,b,t_k,t_l,t_{kl})\,,
\nonumber\\[0.2em]
G_{A_2}(y,r,b,t_k,t_l,t_{kl}) &= G_A(r y,y,b,t_k,t_l,t_{kl})\,.
\end{align}

\section{Calculation of the bare soft function}
\label{sec:baresoftfun}

Having specified the measurement function for two uncorrelated emissions, the calculation of 
the bare soft function defined in \eqref{eq:softfun:definition} proceeds along the lines 
outlined for the correlated-emission contribution in~\cite{Bell:2018oqa}. In the following we 
adopt the notation from that paper and we assume that the Wilson lines are given in the 
fundamental colour representation.

The bare soft function has a double expansion in the dimensional regulator $\eps=(4-d)/2$ and 
the rapidity regulator $\alpha$, which we implement on the level of the phase-space integrals 
via the prescription~\cite{Becher:2011dz}
\begin{equation}
\int d^dp \; \left(\frac{\nu}{n\cdot p + \bar n \cdot p}\right)^\alpha \;  
\delta(p^2) \theta(p^0) \,.
\label{eq:analyticregulator}
\end{equation}
The rapidity regulator is required only for SCET-2 soft functions, and we will introduce an 
alternative version that is compatible with the RRG framework later in Section~\ref{sec:rrg}. 
Up to NNLO the bare soft function can then be written in the form
\begin{align}
S_0(\tau,\nu) = 1 &+ \left(\frac{Z_\alpha\alpha_s}{4\pi}\right) \,(\mu^2 \taubar^2)^\eps \;
(\nu \taubar)^\alpha \, S_R(\eps,\alpha) 
\no\\
& + \left(\frac{Z_\alpha\alpha_s}{4\pi}\right)^2 (\mu^2 \taubar^2)^{2\eps} \; 
\bigg\{(\nu \taubar)^\alpha \,S_{RV}(\eps,\alpha) + (\nu \taubar)^{2\alpha}
\,S_{RR}(\eps,\alpha)\bigg\} + \mathcal{O}(\alpha_s^3)\,,
\label{eq:baresoftfun:expansion}
\end{align}
where $\taubar = \tau e^{\gamma_E}$ and $\alpha_{s}$ is the renormalised strong coupling 
constant in the $\overline{\text{MS}}$ scheme. In~\cite{Bell:2018oqa} we presented the 
calculation of the single real-emission correction $S_R(\eps,\alpha)$, the mixed real-virtual 
inter\-ference $S_{RV}(\eps,\alpha)$ and two out of three colour structures 
($C_{F}C_{A}$, $C_{F}T_{F}n_{f}$) of the double real-emission contribution $S_{RR}(\eps,\alpha)$, 
and the goal of the present paper consists in computing the last missing NNLO ingredient, 
i.e.~the $C_{F}^{2}$ contribution to $S_{RR}(\eps,\alpha)$.

The starting point of our calculation is the representation
\begin{align}
S_{RR}^{(C_F)}(\eps,\alpha) &= \frac{(4\pi e^{\gamma_E} \tau^2)^{-2\eps}\;
\taubar^{-2\alpha} }{(2\pi)^{2d-2}} \, \int d^{d}k \;\, \delta(k^{2}) \,\theta(k^{0})\, 
\int d^{d}l \;\, \delta(l^{2}) \,\theta(l^{0}) 
\no\\[0.2em]
&\quad \times
\frac{|\mathcal{A}_{RR}^{(C_F)}(k,l)|^{2}}
{(n\cdot k + \bar n \cdot k)^\alpha\,(n\cdot l + \bar n \cdot l)^\alpha} \;\, 
\mathcal{M}_2^{unc}(\tau; k,l) \,,
\label{eq:nnlo:doublereal:start}
\end{align}
where -- due to NAE -- the squared matrix element is given by
\begin{align}
|\mathcal{A}_{RR}^{(C_F)}(k,l)|^{2} &= 
\frac{|\mathcal{A}_{R}(k)|^{2}\, |\mathcal{A}_{R}(l)|^{2}}{2}
= \frac{2048 \pi^{4} \,C_{F}^2}{k_+k_-l_+l_-} \,.
\end{align}
From \eqref{eq:nnlo:doublereal:start} it is evident that the calculation reduces to the square
of the NLO soft function if the observable obeys NAE, i.e.~if its double-emission measurement 
function is of the form \eqref{eq:measure:NAE}. We do not assume here, however, that this is the 
case and instead use the more general parametrisation \eqref{eq:measure:NNLO:unc} of the 
measurement function. 

\begin{figure}[t!]
\begin{subfigure}[t]{.32\textwidth}
\includegraphics[width=0.99\textwidth]{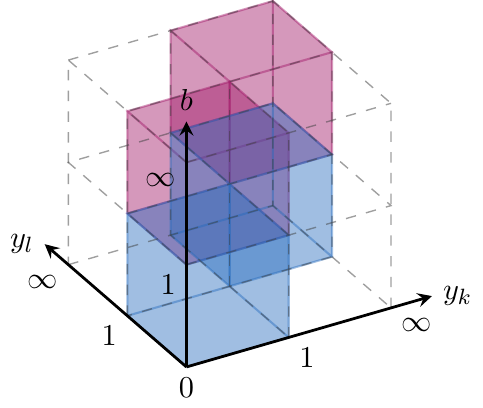}
\caption{$n \leftrightarrow \bar n$ exchange}
\end{subfigure}
\begin{subfigure}[t]{.32\textwidth}
\includegraphics[width=0.99\textwidth]{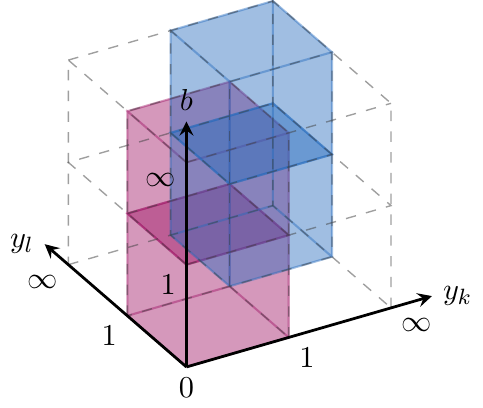}
\caption{$k \leftrightarrow l$ exchange}
\end{subfigure}
\begin{subfigure}[t]{.32\textwidth}
\includegraphics[width=0.99\textwidth]{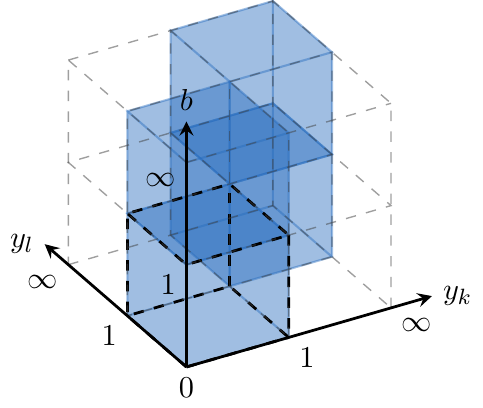}
\caption{Reduced integration region}
\end{subfigure}
\caption{Reduction of the integration domain in the variables $y_k$, $y_l$ and $b$ for
the un\-correlated-emission contribution. Cubes of the same colour correspond to
integration regions which yield the same result according to the stated symmetries. The
second integration region B, complementing the highlighted region A in (c), can be any 
of the white adjacent cubes.  In practice it is most easily recovered from A by inverting 
either $y_k$ or $y_l$. 
\label{fig:symmUC}}
\end{figure}

Starting from \eqref{eq:nnlo:doublereal:start}, we thus switch to the variables introduced in 
\eqref{eq:parametr:nnlo} and \eqref{eq:parametr:nnlo:angles} and perform the observable-independent 
integrations, following~\cite{Bell:2018oqa} for a convenient parametrisation of the angular 
integrals in the $(d-2)$-dimensional transverse plane. In order to map the integration ranges 
in the variables $y_k$, $y_l$ and $b$ onto the unit hypercube, we exploit the fact that the variables 
transform under $n\leftrightarrow\bar n$ exchange as
\begin{equation}
y_k \rightarrow \frac{1}{y_k}\,, \qquad 
y_l \rightarrow \frac{1}{y_l}\,, \qquad 
b \rightarrow b\,, \qquad 
t_k \rightarrow t_k\,, \qquad 
t_l \rightarrow t_l\,, \qquad 
t_{kl} \rightarrow t_{kl}\,,
\end{equation}
whereas the corresponding relations under $k\leftrightarrow l$ exchange are given by
\begin{equation}
y_k \rightarrow y_l\,, \qquad 
y_l \rightarrow y_k\,, \qquad 
b \rightarrow \frac{1}{b}\,, \qquad 
t_k \rightarrow t_l\,, \qquad 
t_l \rightarrow t_k\,, \qquad 
t_{kl} \rightarrow t_{kl}\,.
\end{equation}
Proceeding in analogy to the correlated-emission calculation in~\cite{Bell:2018oqa}, we can 
use these symmetry considerations to map the integration domain onto two independent regions that 
are illustrated in Figure~\ref{fig:symmUC}. In region A, which we take to be the highlighted
dashed blue cube in Figure~\ref{fig:symmUC}(c), the integrand is simply the original integrand 
in which no substitutions are made. The second region B, on the other hand, refers to any of the 
white adjacent cubes in this figure, and it can be most easily recovered from the original integrand 
by inverting either of the variables $y_k$ or $y_l$. 

After performing all of these manipulations, we arrive at the following master formula for the 
calculation of the uncorrelated-emission contribution
\begin{align}
&S_{RR}^{(C_F)}(\eps,\alpha) = 
\frac{128C_F^2\, e^{-2\gamma_E(\eps+\alpha)}\,\Gamma(-4 \eps-2\alpha)}{\pi^{3/2}\, 
\Gamma(- \eps)\,\Gamma(1/2 - \eps)}\;
\int_0^1 \! dy_k \int_0^1 \! dy_l  \int_0^1 \! db 
\int_{0}^1 \!dt_{kl}  \int_{0}^1 \!dt_{l} 
\int_{0}^1 \!dt_{5}'
\no\\[0.2em]
&\qquad\times
b^{-1-2\eps-\alpha}\,(y_k y_l)^{-1+n\eps+(n+1)\alpha/2} \,(1+b)^{4\eps+2\alpha}\; 
\big[(1+y_k)(1+y_l)\big]^{2n\eps+(n-1)\alpha}
\no\\[0.2em]
&\qquad\times
\big(4t_{kl}\bar t_{kl}\big)^{-1/2-\eps}\;
\big(4t_{l}\bar t_{l}\big)^{-1/2-\eps}\;
\big(t_{5}'(2-t_{5}')\big)^{-1-\eps}\; 
\no\\[0.2em]
&\qquad\times
\Big\{G_A(y_k,y_l,b,t_k^+,t_l,t_{kl})^{4\eps+2\alpha}
+ G_B(y_k,y_l,b,t_k^+,t_l,t_{kl})^{4\eps+2\alpha}
+ (t_k^+\to t_k^-)
\Big\}
\label{eq:rr:master}
\end{align}
with 
\begin{equation}
t_{k}^\pm= t_l + t_{kl} - 2 t_{l} t_{kl} \pm 2\sqrt{t_{l}\bar t_{l}t_{kl}\bar t_{kl}}\,(1-t_5') 
\end{equation}
and
\begin{align}
G_A(y_k,y_l,b,t_k,t_l,t_{kl}) &= G(y_k,y_l,b,t_k,t_l,t_{kl})\,,
\nonumber\\[0.2em]
G_B(y_k,y_l,b,t_k,t_l,t_{kl}) &= \begin{dcases}         
\,y_k^{-n} \;G(1/y_k,y_l,b,t_k,t_l,t_{kl}) &\,\text{or} \\
\,y_l^{-n} \;G(y_k,1/y_l,b,t_k,t_l,t_{kl})\,.
\end{dcases}
\label{eq:rr:gb}
\end{align}
In physical terms region A describes the emission of two soft partons into the same hemisphere 
with respect to the collinear axis, whereas region B covers the opposite-hemisphere case. Similar 
to~\cite{Bell:2018oqa}, the expression in region B is not unique, since the symmetry arguments 
only guarantee that the integrals in \eqref{eq:rr:master} are equal, but not necessarily the 
integrands. One is therefore free to derive the functional form of $G_B$ using either of the 
expressions on the right-hand side of \eqref{eq:rr:gb}.

From \eqref{eq:rr:master} we can analyse the divergence structure of the uncorrelated-emission
contribution. For SCET-1 observables with $n\neq 0$, one can set the analytic regulator 
$\alpha$ to zero, and one finds an explicit divergence encoded in $\Gamma(-4 \eps)$
that originates from the analytic integration over the dimensionful variable $q_T$. The integrand
is, moreover, divergent in the limits $b\to 0$, $y_k\to 0$ and $y_l\to 0$ as anticipated 
in Section~\ref{sec:measure}. In addition, there exists a spurious divergence in the limit 
$t_5'\to 0$, which is cancelled by the prefactor $1/\Gamma(-\epsilon)$ as in~\cite{Bell:2018oqa}.
The overall contribution to the bare soft function therefore starts with a $1/\eps^4$ divergence 
for SCET-1 observables.

For SCET-2 soft functions with $n=0$, the analytic regulator cannot be set to zero, since the 
$y_k$ and $y_l$-integrations generate poles in $\alpha$ in this case. As the $\alpha$-expansion 
has to be performed first, the terms $b^{-1-2\eps-\alpha}$ and $\Gamma(-4 \eps-2\alpha)$ introduce  
additional $\eps$-divergences, and they trade $\alpha$-poles for $\eps$-poles in the double 
expansion. The leading divergences in the SCET-2 case are therefore of the form 
$1/(\alpha^{2}\epsilon^{2})$, $1/(\alpha\epsilon^{3})$ and  $1/\epsilon^{4}$.

Finally, we noted towards the end of Section~\ref{sec:measure} that the collinear limits 
$y_k\to0$ and $y_l\to0$ can be ambiguous on the observable level. In order to disentangle 
the joint collinear limit of the emitted partons from the individual ones, we apply a sector 
decomposition strategy in the same-hemisphere contribution and write
\begin{align}
\int_0^1 \! dy_k \,
\int_0^1 \! dy_l \;\;
\mathcal{I}(y_k,y_l)
&= \int_0^1 \! dy \,
\int_0^1 \! dr \;
y\;\bigg\{ \mathcal{I}(y,r y) + \mathcal{I}(r y,y) \bigg\}\,,
\end{align}
where $\mathcal{I}(y_k,y_l)$ symbolically represents the integrand in \eqref{eq:rr:master}, which 
implicitly depends on the other integration variables. This generates two subregions in 
region A with
\begin{align}
G_{A_1}(y,r,b,t_k,t_l,t_{kl}) &= G_A(y,r y,b,t_k,t_l,t_{kl})\,,
\nonumber\\[0.2em]
G_{A_2}(y,r,b,t_k,t_l,t_{kl}) &= G_A(r y,y,b,t_k,t_l,t_{kl})\,.
\label{eq:rr:ga12}
\end{align}
In the numerical implementation of our algorithm we perform a number of additional substitutions
that are designed to improve the numerical convergence. For more details on this technical 
point we refer to Section 6 of~\cite{Bell:2018oqa} and the \softserve~user manual.

\section{Renormalisation}
\label{sec:renormalize}

With the master formula of the uncorrelated-emission contribution at hand, we have assembled all 
ingredients required for the calculation of bare NNLO dijet soft functions. In~\cite{Bell:2018oqa} 
we went one step ahead and extracted the anomalous dimensions and matching corrections that are 
needed for resummations within SCET. To do so, we assumed that the renormalised soft function 
$S = Z_S S_0$ obeys the renormalisation group equation (RGE)
\begin{align}
\frac{\rd}{\rd \ln\mu}  \; S(\tau,\mu)
&= - \frac{1}{n} \,\bigg[ 4 \,\Gamma_{\mathrm{cusp}}(\alpha_s) \, 
\ln(\mu\taubar) 
-2 \gamma^{S}(\alpha_s) \bigg] \; S(\tau,\mu) 
\label{eq:RGE}
\end{align}
for SCET-1 observables, whereas we focused on the collinear anomaly exponent
$\mathcal{F}(\tau,\mu)$ defined via
\begin{align}
S(\tau,\mu,\nu)
&= (\nu^2\taubar^2)^{-\mathcal{F}(\tau,\mu)} \;W^S(\tau,\mu)
\label{eq:ren:scet2:ca}
\end{align}
in the SCET-2 case. The calculations provided in the current paper are fully compatible with 
this setup, and they provide the $C_F^2$ coefficients of the anomalous dimensions 
and matching corrections that were derived in~\cite{Bell:2018oqa} on the basis of NAE.

In this paper we generalise the renormalisation programme in two respects.
First, we consider soft functions that renormalise directly in momentum (or cumulant) space 
rather than Laplace space, which is relevant e.g.~for certain jet-veto observables.
Second, we discuss the renormalisation of SCET-2 soft functions in the RRG 
approach~\cite{Chiu:2012ir}, which is equivalent to the collinear anomaly 
framework from~\cite{Becher:2010tm,Becher:2011pf}, but which requires a specific 
implementation of the rapidity regulator. We will address both of these questions in turn.

\subsection{Cumulant soft functions}
\label{sec:cumulant}

Soft functions for jet-veto observables typically involve measurement functions that are 
formulated in terms of a $\theta$-function, which reflects the fact that the jet veto
provides a cutoff for the phase-space integrations of the soft radiation. Instead of 
the exponential form \eqref{eq:measure:general}, their measurement function can be
expressed as
\begin{align}
\widehat{\mathcal{M}}(\omega;\lbrace k_{i} \rbrace) = 
\theta\big(\omega- \omega(\lbrace k_{i} \rbrace)\,\big)\,,
\end{align}
where $\omega$ is the cutoff variable and the function $\omega(\lbrace k_{i} \rbrace)$
is assumed to obey the same constraints that were listed in detail in Section~\ref{sec:measure}. 

The measurement function of such \emph{cumulant soft functions} can easily be brought 
into the form 
\eqref{eq:measure:general} via a Laplace transformation,
\begin{align}
\int_0^\infty \text{d}\omega\; e^{- \tau \omega} \;
\theta\big(\omega- \omega(\lbrace k_{i} \rbrace)\,\big) 
&= \frac{1}{\tau} \;
\exp\big(-\tau\, \omega(\lbrace k_{i} \rbrace)\,\big)\,.
\end{align}
The factor $1/\tau$ is just a constant for the bare soft function calculation, but it is
relevant for inverting the Laplace transformation. From \eqref{eq:baresoftfun:expansion}
we see that the individual contributions to the soft function come with different powers
of the Laplace variable $\tau$, which can be transformed back to momentum space 
using the relation
\begin{align}
\int_0^\infty \text{d}\omega\; e^{- \tau \omega} \; \omega^m
&= \Gamma(1+m) \; \tau^{-1-m}\,,
\qquad m>-1\,.
\end{align}
Up to NNLO a generic bare cumulant soft function therefore takes the form
\begin{align}
\label{eq:baresoftfun:cumulant}
\widehat{S}_0(\omega,\nu) = 1 &+ \left(\frac{Z_\alpha\alpha_s}{4\pi}\right) 
\left(\frac{\mu^2}{\omega^2}\right)^\eps \left(\frac{\nu}{\omega}\right)^\alpha 
\frac{e^{\gamma_E(2\eps+\alpha)}}{\Gamma(1-2\eps-\alpha)}\; S_R(\eps,\alpha) 
\\[0.2em]
&+ \left(\frac{Z_\alpha\alpha_s}{4\pi}\right)^2 
\left(\frac{\mu^2}{\omega^2}\right)^{2\eps}\,
\bigg\{\left(\frac{\nu}{\omega}\right)^\alpha 
\frac{e^{\gamma_E(4\eps+\alpha)}}{\Gamma(1-4\eps-\alpha)}\; S_{RV}(\eps,\alpha) 
\no\\[0.2em]
& \hspace{40mm} + \left(\frac{\nu}{\omega}\right)^{2\alpha}
\frac{e^{\gamma_E(4\eps+2\alpha)}}{\Gamma(1-4\eps-2\alpha)}\; S_{RR}(\eps,\alpha)\bigg\} 
+ \mathcal{O}(\alpha_s^3)\,,
\no
\end{align}
where the terms $S_i(\eps,\alpha)$ for $i\in\{R,RV,RR\}$ can be calculated with the formulae 
provided in~\cite{Bell:2018oqa} and the present paper, and their prefactors in terms of Euler's
constant and Gamma functions slightly reshuffle the coefficients in the $\eps$ and $\alpha$ 
expansions. They do not modify, however, the divergence structure of the soft function 
since they all expand to $1+\mathcal{O}(\alpha,\eps)$.

We now assume that the RGEs for cumulant soft functions take the same form as \eqref{eq:RGE} and 
the corresponding equation in the SCET-2 case, with the replacement $\bar \tau \to 1/\omega$.
The renormalisation procedure that we developed for Laplace-space soft functions 
in~\cite{Bell:2018oqa} can then be carried over to cumulant soft functions if the
 prefactors in \eqref{eq:baresoftfun:cumulant} are included. As we will explain later in
Section~\ref{sec:softserve}, \softserveone~contains a script for the renormalisation of cumulant
soft functions which applies these modifications and which takes the correct error propagation 
into account.

\subsection{Rapidity renormalisation group}
\label{sec:rrg}

The collinear anomaly~\cite{Becher:2010tm,Becher:2011pf} and the RRG~\cite{Chiu:2012ir}
provide two equivalent frameworks for the renormalisation of SCET-2 soft functions. 
In the latter the soft function is renormalised via multiplication with a Z-factor, 
$S = Z_S S_0$, that absorbs the divergences both in the dimensional regulator $\eps$ and the 
rapidity regulator $\alpha$. The renormalised soft function is furthermore 
assumed to satisfy the RRG equation 
\begin{align}
\frac{\rd}{\rd \ln\nu}  \; S(\tau,\mu,\nu)
&= \bigg[ 4 \,A_\Gamma(\mu_s,\mu) 
-2 \gamma_{\nu}^S(\tau,\mu_s) \bigg] \; S(\tau,\mu,\nu) \,,
\label{eq:RRG:nuRGE}
\end{align}
where $A_\Gamma(\mu_1, \mu_2)$ is an RG kernel that was given explicitly in~\cite{Bell:2018oqa},
and the $\nu$-anomalous dimension can be identified with the collinear anomaly exponent
defined in \eqref{eq:ren:scet2:ca} via
\begin{align}
\gamma_{\nu}^S(\tau,\mu_s) &=\mathcal{F}(\tau,\mu_s)\,.
\end{align}
In the RRG approach the renormalised soft function is in addition supposed to obey a RGE in 
the scale $\mu$,
\begin{align}
\frac{\rd}{\rd \ln\mu}  \; S(\tau,\mu,\nu)
&= \bigg[ 4 \,\Gamma_{\mathrm{cusp}}(\alpha_s) \, \ln(\mu\taubar) 
- 4 \,\Gamma_{\mathrm{cusp}}(\alpha_s) \, \ln(\nu\taubar) 
-2 \gamma^{S}_\mu(\alpha_s) \bigg] \; S(\tau,\mu,\nu) \,,
\label{eq:RRG:muRGE}
\end{align}
whereas the corresponding quantity in the collinear anomaly framework -- the soft remainder
function $W^S(\tau,\mu)$ in \eqref{eq:ren:scet2:ca} -- does not obey a simple RGE without its 
collinear counterpart. The RRG therefore makes stronger assumptions than the collinear anomaly 
framework, and we argued in~\cite{Bell:2018oqa} that the RGE \eqref{eq:RRG:muRGE} only holds if 
the rapidity regulator is implemented on the level of connected webs -- a necessary requirement
for the consistency of the RRG approach that was not formulated so clearly in the original 
literature.\footnote{Connected webs were discussed in~\cite{Chiu:2012ir} only in the context 
of gauge invariance and NAE, but it has not been stated explicitly in that paper that the RGE 
\eqref{eq:RRG:muRGE} looses its validity if the rapidity regulator is not implemented on
the level of connected webs.}

As we implement the rapidity regulator via the prescription \eqref{eq:analyticregulator} 
for individual emissions, our default setup is not suited for the RRG approach. In other words the 
\mbox{$\alpha^0$-pieces} calculated with \softserveonine~cannot be renormalised in a way that is 
consistent with \eqref{eq:RRG:muRGE} (as the problem does not affect the $1/\alpha$ poles,
all results presented in~\cite{Bell:2018vaa,Bell:2018oqa,Bell:2018jvf} are never\-theless correct).
In \softserveone~we remedy this point and implement an alter\-native prescription that fulfils the 
requirements of the RRG approach. To do so, we add a factor $w^2$ to 
\eqref{eq:analyticregulator}, where $w$ is a bookkeeping parameter that fulfils the
RRG equation $\rd w/\rd \ln\nu = -\alpha w/2$~\cite{Chiu:2012ir}, and we implement the  
rapidity regulator for double correlated emissions via
\begin{equation}
w^2\,\int d^dk \int d^dl \; \left(\frac{\nu}{k_+ + k_- + l_+ + l_-}\right)^\alpha \;  
\delta(k^2) \theta(k^0) \;
\delta(l^2) \theta(l^0)
\label{eq:analyticregulator:RRG:correlated}
\end{equation}
rather than
\begin{equation}
w^4\int d^dk \; \left(\frac{\nu}{k_+ + k_-}\right)^\alpha \;  
\delta(k^2) \theta(k^0) \;
\int d^dl \; \left(\frac{\nu}{l_+ + l_-}\right)^\alpha \;  
\delta(l^2) \theta(l^0)\,,
\label{eq:analyticregulator:RRG:correlated:not}
\end{equation}
whereas the remaining contributions to the bare soft function are not changed, except for  
trivial factors of $w$. 

We will address the technical aspects of the \softserve~implementation in the following 
section, and show here how to extract the two-loop anomalous dimensions and matching corrections 
from the bare soft function in the RRG setup. To do so, we start from 
\begin{align}
\label{eq:ren:scet2bare}
S_0(\tau,\nu) &= 1 + \left(\frac{Z_\alpha\alpha_s}{4\pi}\right) \,w^2
\,(\mu^2 \taubar^2)^\eps \;(\nu \taubar)^\alpha 
\bigg\{ \frac{1}{\alpha} \bigg(
\frac{x^1_1}{\eps}  + x^1_0 + x^1_{-1}\,\eps + x^1_{-2}\,\eps^2 + x^1_{-3}\,\eps^3\bigg) 
\\[0.2em]
& + 
\frac{x^0_2}{\eps^2}  + \frac{x^0_1}{\eps} + x^0_0 + x^0_{-1}\,\eps + x^0_{-2}\,\eps^2
+\alpha \bigg(
\frac{x^{-1}_3}{\eps^3} + \frac{x^{-1}_2}{\eps^2} + \frac{x^{-1}_1}{\eps}  + x^{-1}_0 + x^{-1}_{-1}\,\eps \bigg) 
\no\\[0.2em]
& +\mathcal{O}\Big(\frac{\eps^4}{\alpha},\eps^3,\alpha\eps^2,\alpha^2\Big)\bigg\}
+ \left(\frac{Z_\alpha\alpha_s}{4\pi}\right)^2 (\mu^2 \taubar^2)^{2\eps} \; \bigg\{ 
w^4 \,(\nu \taubar)^{2\alpha} \,\bigg[ \frac{1}{\alpha^2} \bigg(
\frac{y^2_2}{\eps^2} + \frac{y^2_1}{\eps}  + y^2_0 \bigg) 
\no\\[0.2em]
& + 
\frac{1}{\alpha} \bigg(
\frac{y^1_3}{\eps^3}  + \frac{y^1_2}{\eps^2}  + \frac{y^1_1}{\eps}  + y^1_0 \bigg) + \frac{y^0_4}{\eps^4} + \frac{y^0_3}{\eps^3}  + \frac{y^0_2}{\eps^2} + \frac{y^0_1}{\eps} + y^0_0 
+\mathcal{O}\Big(\frac{\eps}{\alpha^2},\frac{\eps}{\alpha},\eps,\alpha\Big)
\bigg]\no\\[0.2em]
& + w^2 \,(\nu \taubar)^{\alpha} \bigg[ 
\frac{1}{\alpha} \bigg(
\frac{z^1_3}{\eps^3}  +\frac{z^1_2}{\eps^2}  +\frac{z^1_1}{\eps}  + z^1_0\bigg) + \frac{z^0_4}{\eps^4} + \frac{z^0_3}{\eps^3}  + \frac{z^0_2}{\eps^2} + \frac{z^0_1}{\eps} + z^0_0
+\mathcal{O}\Big(\frac{\eps}{\alpha},\eps,\alpha\Big)\bigg] \bigg\},
\no
\end{align}
where the only difference with respect to~\cite{Bell:2018oqa} consists in the presence of the 
bookkeeping parameter $w$. Due to \eqref{eq:analyticregulator:RRG:correlated} the 
correlated-emission contribution is, moreover, now contained in the $z^i_j$ coefficients along
with the real-virtual interference term. The single real-emission and uncorrelated double-emission 
contributions constitute the $x^i_j$ and $y^i_j$ coefficients, respectively, as before.
The coefficients $x^i_j$ are thus proportional to the colour factor $C_F$, the $y^i_j$ to $C_F^2$, 
and the $z^i_j$ consist of two contributions with colour factors $C_F T_f n_f$ and $C_F C_A$.

We now expand the anomalous dimensions to two-loop order,
\begin{align}
\Gamma_{\mathrm{cusp}}(\alpha_s) &= \left(\frac{\alpha_{s}}{4 \pi} \right) \Gamma_{0}
+ \left(\frac{\alpha_{s}}{4 \pi} \right)^{2} \Gamma_{1}\,, 
\\[0.2em]
\gamma_\mu^{S}(\alpha_s)&= \left(\frac{\alpha_{s}}{4 \pi} \right) \gamma^{S}_{\mu,0}
+\left(\frac{\alpha_{s}}{4 \pi} \right)^{2} \gamma^{S}_{\mu,1}\,, 
\no\\[0.2em]
\gamma_\nu^{S}(\tau,\mu) &= 
\left( \frac{\alpha_s}{4 \pi} \right) 
\Big\{ 2\Gamma_0 L_\mu 
+ \gamma^{S}_{\nu,0} \Big\}
+\left( \frac{\alpha_s}{4 \pi} \right)^2 
\Big\{ 2 \beta_0\Gamma_0 L_\mu^2 
 + 2 \left( \Gamma_1 + \beta_0  \gamma^{S}_{\nu,0} \right) L_\mu 
+ \gamma^{S}_{\nu,1} \Big\},
 \no
\end{align}
where $L_\mu=\ln(\mu\taubar)$ and the coefficients $\gamma^{S}_{\nu,i}$ correspond to the 
$d_{i+1}$ in the collinear anomaly language of~\cite{Bell:2018oqa}. Using 
$Z_\alpha = 1-\beta_0\alpha_s/(4\pi\eps) + \mathcal{O}(\alpha_s^2)$,
we can solve the RGEs \eqref{eq:RRG:nuRGE} and \eqref{eq:RRG:muRGE} for the soft function
and the corresponding equations for the Z-factor $Z_{S}=S/S_0$  explicitly. In order to avoid cross 
terms from higher orders, the latter is conveniently determined via its logarithm, which in the 
$\overline{\text{MS}}$ scheme takes the form 
\begin{align} 
& \ln Z_{S} = \left( \frac{\alpha_s}{4 \pi} \right) w^2\,
\bigg\{ \frac{2\Gamma_0}{\alpha \eps} +\frac{4\Gamma_0L_\mu +2\gamma_{\nu,0}^S}{\alpha} 
-\frac{\Gamma_0}{\eps^2} +\frac{\gamma_{\mu,0}^S-2\Gamma_0(L_\mu-L_\nu)}{\eps}\bigg\}
\no\\[0.2em]
&\quad 
+\left( \frac{\alpha_s}{4 \pi} \right)^2 w^2\,
\bigg\{ -\frac{\beta_0 \Gamma_0}{\alpha\eps^2} + \frac{\Gamma_1}{\alpha\eps} + 
\Big(4 \beta_0 \Gamma_0 L_\mu^2 + 4 (\Gamma_1 + \beta_0 \gamma_{\nu,0}^S) L_\mu 
+ 2 (\gamma_{\nu,1}^S)_C + w^2 (\gamma_{\nu,1}^S)_U\Big)\frac{1}{\alpha} 
\no\\[0.2em]
&\hspace{28mm}
+ \frac{3 \beta_0 \Gamma_0}{4\eps^3} - \frac14 \Big(\Gamma_1 + 2 \beta_0 \gamma_{\mu,0}^S 
   - 4 \beta_0 \Gamma_0 (L_\mu - L_\nu)\Big)\frac{1}{\eps^2} 
\no\\[0.2em]
&\hspace{28mm}   
   - \frac12 \Big(2 \Gamma_1 (L_\mu - L_\nu) 
   - (\gamma_{\mu,1}^S)_C - w^2 (\gamma_{\mu,1}^S)_U\Big)\frac{1}{\eps}\bigg\}\,,
\label{eq:logZS}
\end{align}
where $L_\nu=\ln(\nu\taubar)$ and we have split the correlated and uncorrelated-emission
contributions to the two-loop anomalous dimensions $\gamma^{S}_{\mu,1}$ and 
$\gamma^{S}_{\nu,1}$ since -- according to \eqref{eq:ren:scet2bare} -- they come with different 
powers of the bookkeeping parameter $w$. For the renormalised soft function, we obtain up to the 
considered two-loop order
\begin{align}
& \ln S(\tau,\mu,\nu) = 
\left( \frac{\alpha_s}{4 \pi} \right) 
\bigg\{ 2\Gamma_0 L_\mu^2 - 4\Gamma_0 L_\mu L_\nu  
- 2\gamma^{S}_{\mu,0} L_\mu 
- 2\gamma^{S}_{\nu,0} L_\nu + c_1^S \bigg\}
\no\\[0.2em]
&\quad 
+\left( \frac{\alpha_s}{4 \pi} \right)^2 
\bigg\{ \frac{4}{3} \beta_0 \Gamma_0 L_\mu^3 
- 4 \beta_0 \Gamma_0 L_\mu^2 L_\nu 
+2 \Big(\Gamma_1 - \beta_0 \gamma_{\mu,0}^S\Big) L_\mu^2 
- 4 \Big(\Gamma_1 + \beta_0 \gamma_{\nu,0}^S\Big) L_\mu L_\nu 
\no\\[0.2em]
&\hspace{22mm}
- 2 \Big(\gamma_{\mu,1}^S - \beta_0 c_1^S\Big) L_\mu - 
 2 \gamma_{\nu,1}^S  L_\nu
+ c_2^S - \frac12 (c_1^S)^2
\bigg\}\,,
\label{eq:logSren}
\end{align}
where we have set $w=1$.  As the cusp anomalous dimension and the beta function are known 
to the required order,
\begin{equation} 
\Gamma_0 = 4 C_F\,,
\quad
\Gamma_1 = 4 C_F
\left\{ \left(\frac{67}{9}-\frac{\pi^2}{3}\right) C_A
- \frac{20}{9} T_F n_f \right\},
\quad
\beta_0 = \frac{11}{3} C_A - \frac43T_F n_f\,,
\end{equation}
the higher poles in the product of the Z-factor and the bare soft function provide checks
of our calculation, whereas the coefficients of the $1/\alpha$ and $1/\eps$ poles determine 
the rapidity anomalous dimension $\gamma^{S}_{\nu}$ and the $\mu$-anomalous dimension 
$\gamma^{S}_{\mu}$, respectively. In terms of the coefficients introduced 
in \eqref{eq:ren:scet2bare}, we obtain
\begin{align}
\gamma^{S}_{\nu,0}&=-\frac{x^1_0}{2}\,,
 \no\\
\gamma^{S}_{\nu,1} &= -y^1_0 - \frac{z^1_0}{2}  + 
 x^0_{-1} x^1_1 +  x^0_{0} x^1_0
+  x^0_{1} x^1_{-1} +  x^0_{2} x^1_{-2} + \frac{\beta_0 x^1_{-1}}{2}\,,
\end{align}
which is precisely the relation we found for the collinear anomaly exponent 
in~\cite{Bell:2018oqa}. The non-logarithmic terms of the renormalised soft function are,
on the other hand, in the RRG framework given by
\begin{align}
c_1^S&=x^0_0\,,
 \\[0.2em] 
c_2^S&= y^0_0 + z^0_0 - x^0_2 x^0_{-2} -(x^0_1+ \beta_0) x^0_{-1} 
-x^1_1 x^{-1}_{-1} - x^1_0 x^{-1}_0 - x^1_{-1} x^{-1}_1 
- x^1_{-2} x^{-1}_2 - x^1_{-3} x^{-1}_3\,,\no
\end{align}
whereas one can show that the $\mu$-anomalous dimension is unphysical for SCET-2 soft functions 
since it drops out in the final expressions once the soft and collinear RG kernels are combined. 
Following the procedure outlined in~\cite{Bell:2018vaa}, we can actually prove that the 
$\mu$-anomalous dimension is a universal number in our setup, i.e.~it is independent of the 
observable given by
\begin{align} 
\gamma^{S}_{\mu,0}=0\,,
\quad
\gamma^{S}_{\mu,1}= 
\bigg\{ \frac{224}{27} - \frac{4\pi^2}{9} \bigg\} 
\, C_F T_F n_f
\,+\, \bigg\{ - \frac{808}{27} + \frac{11\pi^2}{9} 
+ 28\zeta_3\bigg\} \,C_F C_A\,.
\end{align}
Rather than extracting this quantity from the coefficient of the $1/\eps$ pole, we therefore turn 
the argument around and use these numbers in \softserve~to check if the singularities cancel out 
as predicted by the RRG framework.

The discussion of cumulant soft functions from the previous section applies identically to the 
RRG setup, with the sole exception that the correlated-emission contribution 
in \eqref{eq:baresoftfun:cumulant} comes with a prefactor 
$e^{\gamma_E(4\eps+\alpha)}/\Gamma(1-4\eps-\alpha)$ 
rather than
$e^{\gamma_E(4\eps+2\alpha)}/\Gamma(1-4\eps-2\alpha)$ 
because of \eqref{eq:analyticregulator:RRG:correlated}. Once again, \softserveone~provides a script 
that takes these modifications into account.

\section{Extending the \softserve~distribution}
\label{sec:softserve}

The central new element of \softserveone~is the direct calculation of the uncorrelated-emission 
contribution, whereas \softserveonine~reconstructs this term from the NLO correction, assuming that 
the observable is consistent with NAE. For the \softserve~user, this means that calling 
\texttt{make all} -- or calling \texttt{make} without target -- now generates executables for all 
colour structures, and the target list is supplemented with the \texttt{uncorrelated}, \texttt{CFA} 
and \texttt{CFB} targets. The latter correspond to the two contributions from regions A and B in 
\eqref{eq:rr:master}, and \texttt{uncorrelated} refers to them as a pair. For observables obeying 
NAE, the \texttt{correlated} target now provides all the required input, skipping the $C_F^2$ 
contributions.\footnote{In version 0.9 \texttt{correlated} was synonymous to \texttt{all}, or no 
target at all.} 

In addition we implemented the new features discussed in Section \ref{sec:renormalize} concerning 
cumulant soft functions and the RRG. Apart from the existing script for the renormalisation of 
Laplace-space soft functions (\texttt{laprenorm}), there now also exists a script for the 
renormalisation of cumulant soft functions (\texttt{momrenorm}) that applies the changes 
discussed in Section \ref{sec:cumulant}. Both scripts come in two versions designed for observables 
that obey NAE (postfix \texttt{NAE}) and those that violate NAE (no postfix). The latter require 
the full set of results files, whereas the former do not need the \texttt{CFA} and \texttt{CFB} 
results --  they reconstruct the $C_F^2$ contribution directly from the NLO result. Execution and 
summary scripts to run and refine the results now also exist in two versions for observables 
that obey/violate NAE, similarly postfixed. To prevent accidentally calling non-\texttt{NAE} 
scripts on results that are derived assuming NAE, some safeguards are implemented. 

Moreover, the SCET-2 executables can now be generated with a rapidity regulator that is 
compatible with the RRG approach. As discussed in Section \ref{sec:rrg}, this requires that 
one implements the regulator on the level of connected webs rather than individual emissions. 
At NNLO the only difference arises in the correlated-emission contribution for which the 
regulator is implemented via \eqref{eq:analyticregulator:RRG:correlated} rather than
\eqref{eq:analyticregulator:RRG:correlated:not}. This feature is switched off by default, but 
it can be used by setting a nonzero \texttt{RRG} variable during the \texttt{make} call. In 
other words, to generate e.g.~the $C_{F}T_{F}n_{f}$ colour structure binary for some observable 
using the RRG regulator, one calls \texttt{make NF RRG=1}. In the SCET-2 branch, there are 
scripts to summarise (\texttt{sftsrvres}), renormalise (\texttt{laprenorm} or \texttt{momrenorm}) 
and to account for Fourier phases (\texttt{fourierconvert}) that use the results derived with the 
new regulator, and they are all postfixed \texttt{RRG}. These scripts of course also exist for 
observables that obey NAE, and they then simply carry both postfixes like \texttt{laprenormNAERRG}. 
Again, safeguards to avoid calling \texttt{RRG} scripts on results that were derived with the 
default rapidity regulator and vice-versa are implemented.

Finally, we added the formulae derived in~\cite{Bell:2018vaa} that allow for a direct calculation 
of the soft anomalous dimensions and collinear anomaly exponents without having to calculate the 
complete bare soft function. As the \softserve~input differs slightly from the conventions 
of~\cite{Bell:2018vaa}, we rederived these formulae in a form that is suitable for \softserve~and 
summarise the corresponding expressions in Appendix \ref{app:AD}.  To access these formulae the 
user must call \texttt{make} with targets \texttt{ADLap} or \texttt{ADMom}, which generates the 
respective executables for Laplace-space and cumulant soft functions. These executables then 
reside in the \texttt{Executables} folder and must be called manually. While they allow for a 
fast evaluation of the anomalous dimension/anomaly exponent, we do not recommend using them for a 
precision determination since they are numerically less robust. Observables 
which exhibit features that reduce numerical accuracy, like integrable divergences, slow them down 
disproportionately. In addition, the term \eqref{eq:CF2:deltagamma1}, which is conjectured to vanish 
for all observables, happens to sometimes be numerically unstable due to the peculiar structure 
in its last line. For observables for which this expression is non-trivial, the integration 
converges comparatively slowly.

\section{Results}
\label{sec:results}

We are now in a position to use \softserveone~to compute NNLO dijet soft functions for various 
$e^+e^-$ event shapes and hadron-collider observables. As in~\cite{Bell:2018oqa}, we present our 
results for SCET-1 soft functions in the form
\begin{align} 
\gamma_0^S &= \gamma_0^{C_F} \,C_F\,,
\no\\[0.2em]
\gamma_1^S &= \gamma_1^{C_A} \,C_F C_A 
+ \gamma_1^{n_f} \,C_F T_F n_f
+ \gamma_1^{C_F} \,C_F^2\,,
\no\\[0.2em]
c_1^S &= c_1^{C_F} \,C_F\,,
\no\\[0.2em]
c_2^S &= c_2^{C_A} \,C_F C_A 
+ c_2^{n_f} \,C_F T_F n_f
+ c_2^{C_F} \,C_F^2\,,
\label{eq:scet-1:results}
\end{align}
where the coefficients $\gamma^S_i$ of the soft anomalous dimension and the finite terms $c_i^S$ 
of the renormalised soft function refer to the conventions introduced in 
Section 4.1~of~\cite{Bell:2018oqa}. In contrast to that work, we now use \softserve~to calculate 
the $\gamma_1^{C_F}$ and $c_2^{C_F}$ numbers, which were derived in~\cite{Bell:2018oqa} on the 
basis of NAE.

For SCET-2 soft functions we quote our numbers in the RRG notation of Section \ref{sec:rrg}. The 
relevant resummation ingredients are in this case the coefficients $\gamma_{\nu,i}^S$ of the 
rapidity anomalous dimension and the finite terms $c_i^S$ of the RRG renormalised soft function, 
which we decompose analogously to \eqref{eq:scet-1:results} according to their colour structures. 
Whereas the former are equivalent to the anomaly coefficients $d_{i+1}$ used in~\cite{Bell:2018oqa}, 
the latter are not well defined in the collinear anomaly framework and were therefore not given 
in~\cite{Bell:2018oqa}. As explained in Section \ref{sec:rrg}, the \mbox{$\mu$-anomalous} 
dimension $\gamma_{\mu}^S$ is, moreover, unphysical for SCET-2 soft functions and will therefore be 
disregarded in the following.    

Similar to~\cite{Bell:2018oqa}, \softserveone~comes with a number of template files that can be 
used to rederive the numbers quoted in this section. For most of the observables the runtime of 
the uncorrelated-emission contribution turns out to be comparable to the correlated-emission 
calculation, which can of course be tailored to the specific needs of the user by adjusting the 
respective Cuba settings.\footnote{As in~\cite{Bell:2018oqa}, the numbers presented in this section 
were produced with the precision setting, while the plots were produced with the standard setting.} 
Although the focus of the present paper is on NAE-violating observables, we first consider a few 
observables that respect NAE, since this allows us to test the new algorithm and to gauge the 
accuracy of our numerical predictions. We then switch to some exemplary NAE-violating soft 
functions in a second step.

\subsection{Observables that obey NAE}

For all observables in this section NAE implies $\gamma_1^{C_F}=0$ and $c_2^{C_F}=1/2(c_1^{C_F})^2$
for SCET-1 soft functions, and similarly $\gamma_{\nu,1}^{C_F}=0$ and $c_2^{C_F}=1/2(c_1^{C_F})^2$
in the SCET-2 case.


\subsubsection*{C-parameter}

We first consider the C-parameter event shape, which was one of the template observables we studied 
in~\cite{Bell:2018oqa}. The only new element required for the uncorrelated-emission contribution is 
the function\footnote{As in~\cite{Bell:2018oqa} we suppress the angular variables in the arguments 
of the measure\-ment function if the observable does not depend on any of  these angles.}
\begin{align}
G(y_k,y_l,b)&=
\frac{1}{(1+y_k)(1+y_l)}
\end{align}
defined in \eqref{eq:measure:NNLO:unc}, which can be translated into the relevant input functions 
$G_{A_1}$, $G_{A_2}$ and $G_{B}$ using the relations \eqref{eq:rr:gb} and \eqref{eq:rr:ga12}. We 
then find using \softserveone
\begin{alignat}{6}
\gamma_0^{C_F}&=  1\cdot 10^{-10} \pm 2\cdot 10^{-7}\, 
&&\quad[0]\,,
&\qquad\quad
c_1^{C_F}&= -3.28987\pm 9\cdot 10^{-7}\, 
&&\quad[-3.28987]\,, 
\no\\[0.2em]
\gamma_1^{C_A}&= 15.7940(10)\, 
&&\quad[15.7945]\,, 
& c_2^{C_A}&= -57.9814(35)
&&\quad[-57.9757]\,, 
\no\\[0.2em]
\gamma_1^{n_f}&= 3.90983(14)\, 
&&\quad[3.90981]\,, 
&c_2^{n_f}&= 43.8181(4)\, 
&&\quad[43.8182]\,,
\no\\[0.2em]
\gamma_1^{C_F}&= -0.0004(24)
&&\quad[0]\,, 
&c_2^{C_F}&= 5.41178(592)
&&\quad[5.41162]\,,
\end{alignat}
which is in excellent agreement with the analytic results from~\cite{Hoang:2014wka,Bell:2018oqa} 
shown in the square brackets.


\subsubsection*{W-production at large transverse momentum}

We next consider the soft function for $W$-production at large transverse momentum which we also 
discussed in detail in~\cite{Bell:2018oqa}. We now have
\begin{align}
G(y_k,y_l,b,t_k,t_l,t_{kl})&\!=\!
\frac{b(1+y_l)(1 + y_k - 2\sqrt{y_k}(1-2t_k))}{(1+b)}
+\frac{(1+y_k)(1 + y_l - 2\sqrt{y_l}(1-2t_l))}{(1+b)}
\end{align}
and obtain
\begin{alignat}{6}
\gamma_0^{C_F}&=  -1\cdot 10^{-9} \pm 2\cdot 10^{-6}\, 
&&\quad[0]\,,
&\qquad\quad
c_1^{C_F}&= 9.86960(2)\, 
&&\quad[9.86960]\,, 
\no\\[0.2em]
\gamma_1^{C_A}&= 15.7945(24)\, 
&&\quad[15.7945]\,, 
& c_2^{C_A}&=-2.64324(890)
&&\quad[-2.65010]\,, 
\no\\[0.2em]
\gamma_1^{n_f}&= 3.90987(22)\, 
&&\quad[3.90981]\,, 
&c_2^{n_f}&= -25.3069(10)\, 
&&\quad[-25.3073]\,,
\no\\[0.2em]
\gamma_1^{C_F}&= -1\cdot 10^{-7} \pm 0.003
&&\quad[0]\,, 
&c_2^{C_F}&= 48.7050(96) 
&&\quad[48.7045]\,,
\end{alignat}
which is again in perfect agreement with the analytic results from~\cite{Becher:2012za}.


\subsubsection*{Jet broadening}

In order to illustrate the new RRG routine of \softserve, we consider the SCET-2 event-shape 
variable jet broadening. As in~\cite{Bell:2018oqa} we consider a recoil-free definition here and 
refer to that paper for more details on the observable. The relevant input for the 
uncorrelated-emission contribution is then given by
\begin{align}
G(y_k,y_l,b)&=
\frac{1}{2}\,,
\end{align}
which yields
\begin{alignat}{6}
\gamma_{\nu,0}^{C_F}&= -5.54518(1)   
&&\quad[-5.54518]\,,
&\qquad\quad
c_1^{C_F}&= -20.2930(1) 
&&\quad[-20.2930]\,,
\no\\[0.2em]
\gamma_{\nu,1}^{C_A}&= 7.03652(110) 
&&\quad[7.03605]\,,
& c_2^{C_A}&= -56.6537(21)\,,
&&
\no\\[0.2em]
\gamma_{\nu,1}^{n_f}&= -11.5393(1)
&&\quad[-11.5393]\,,
&c_2^{n_f}&= 24.1971(3)\,, 
&&
\no\\[0.2em]
\gamma_{\nu,1}^{C_F}&= -0.00001(163) 
&&\quad[0]\,,
&c_2^{C_F}&=  205.902(5)  
&&\quad[205.902]\,.
\end{alignat}
For the rapidity anomalous dimension, this agrees with the expressions found 
in~\cite{Becher:2012qc}, and the one-loop matching coefficient $c_1^{C_F}= -8 \ln^2 2-5 \pi^2/3$ 
can be extracted from that paper as well. Our results for the two-loop coefficients $c_2^{C_A}$ 
and $c_2^{n_f}$ are, on the other hand, new. 


\subsubsection*{Transverse-momentum resummation}

We finally examine the soft function for transverse-momentum resummation in Drell-Yan production, 
which is an example of a Fourier-space rather than a Laplace-space soft function. As argued 
in~\cite{Bell:2018oqa}, these can be computed with \softserve~by using the absolute value of 
the naive measurement function, which in the specific case of transverse-momentum resummation 
is given by
\begin{align}
G(y_k,y_l,b,t_k,t_l,t_{kl})&=
\frac{2}{1+b}\;\big|b (1-2t_k) + 1-2t_l\big|\,.
\end{align}
Running the \texttt{fourierconvertRRG} script before renormalisation, we then obtain
\begin{alignat}{6}
\gamma_{\nu,0}^{C_F}&= 1\cdot 10^{-9}\pm 2\cdot 10^{-6}
&&\quad[0]\,,
&\qquad\quad
c_1^{C_F}&= -3.2899(1) 
&&\quad[-3.2899]\,,
\no\\[0.2em]
\gamma_{\nu,1}^{C_A}&= -3.7407(94)  
&&\quad[-3.7317]\,,
& c_2^{C_A}&= -16.749(169)  
&&\quad[-16.507]\,,
\no\\[0.2em]
\gamma_{\nu,1}^{n_f}&= -8.2963(20) 
&&\quad[-8.2963]\,,
&c_2^{n_f}&= 10.338(27)  
&&\quad[10.347]\,,
\no\\[0.2em]
\gamma_{\nu,1}^{C_F}&= -0.0322(281) 
&&\quad[0]\,,
&c_2^{C_F}&=  5.1718(3987)  
&&\quad[5.4116]\,.
\end{alignat}
While we already calculated the rapidity anomalous dimension for this observable 
in~\cite{Bell:2018oqa}, we did not have access to the finite terms in the RRG framework at the 
time, which are however known analytically from the calculation in~\cite{Luebbert:2016itl}. 
Our \softserve~numbers compare well to these results, although we observe a slightly reduced 
accuracy in comparison to the prior examples, which is due to integrable divergences in the 
bulk of the integration region as well as the required Fourier shuffle, which mixes coefficients 
and adds up the corresponding errors. The agreement is, however, still acceptable.

\subsection{Observables that violate NAE}

Having established that \softserveone~satisfactorily reproduces known results for sample NAE 
observables, we now turn to soft functions that do not respect the NAE theorem and which 
require an independent calculation of the uncorrelated-emission contribution. Whereas we already 
presented our results for the corresponding anomalous dimensions in~\cite{Bell:2018vaa,Bell:2018jvf},
we compute the  matching coefficients in this work for the first time.


\subsubsection*{Rapidity-dependent jet vetoes}

\begin{figure}[t]
\centering
\includegraphics[width=0.32\textwidth]{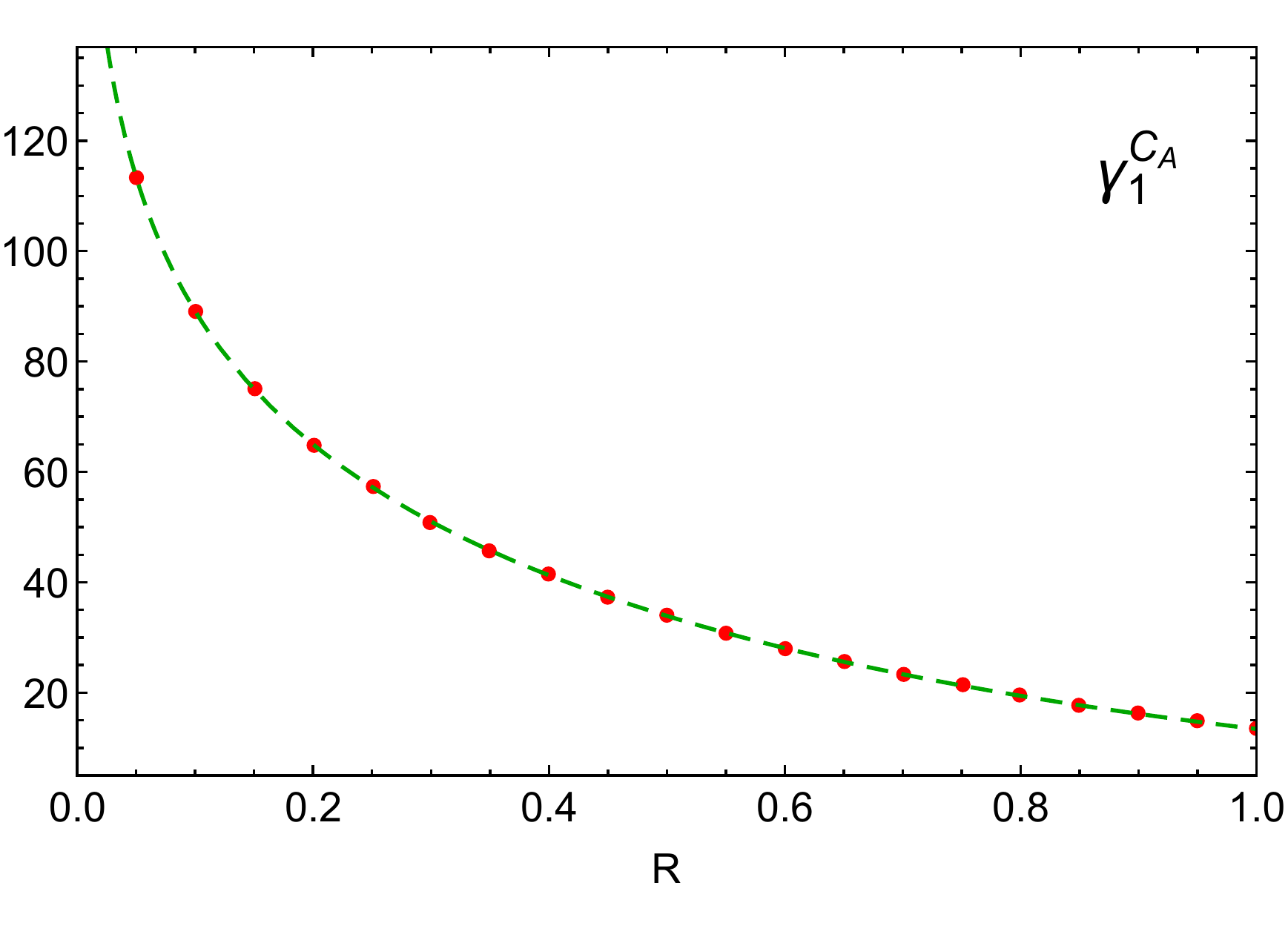}
\includegraphics[width=0.32\textwidth]{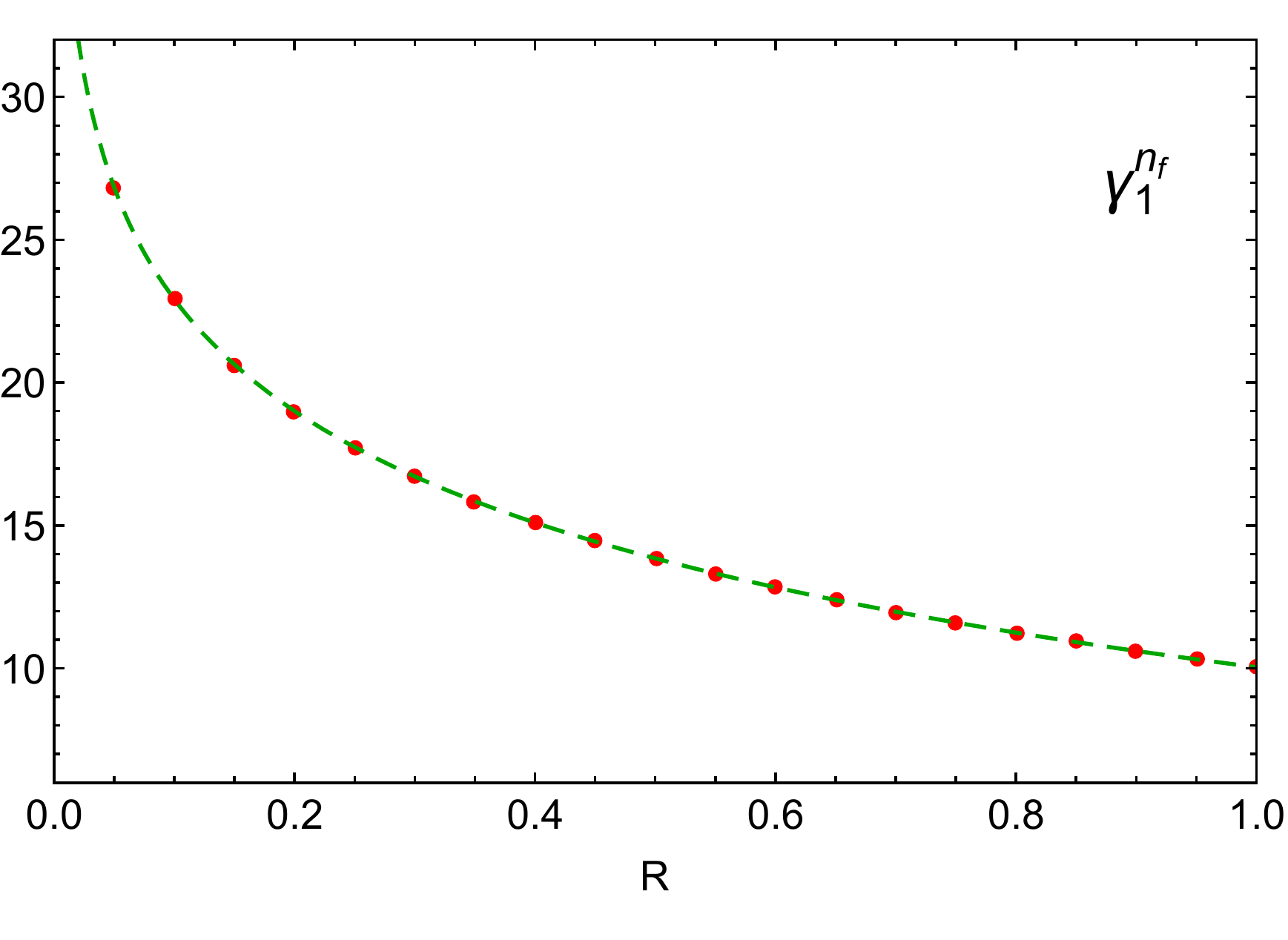}
\includegraphics[width=0.32\textwidth]{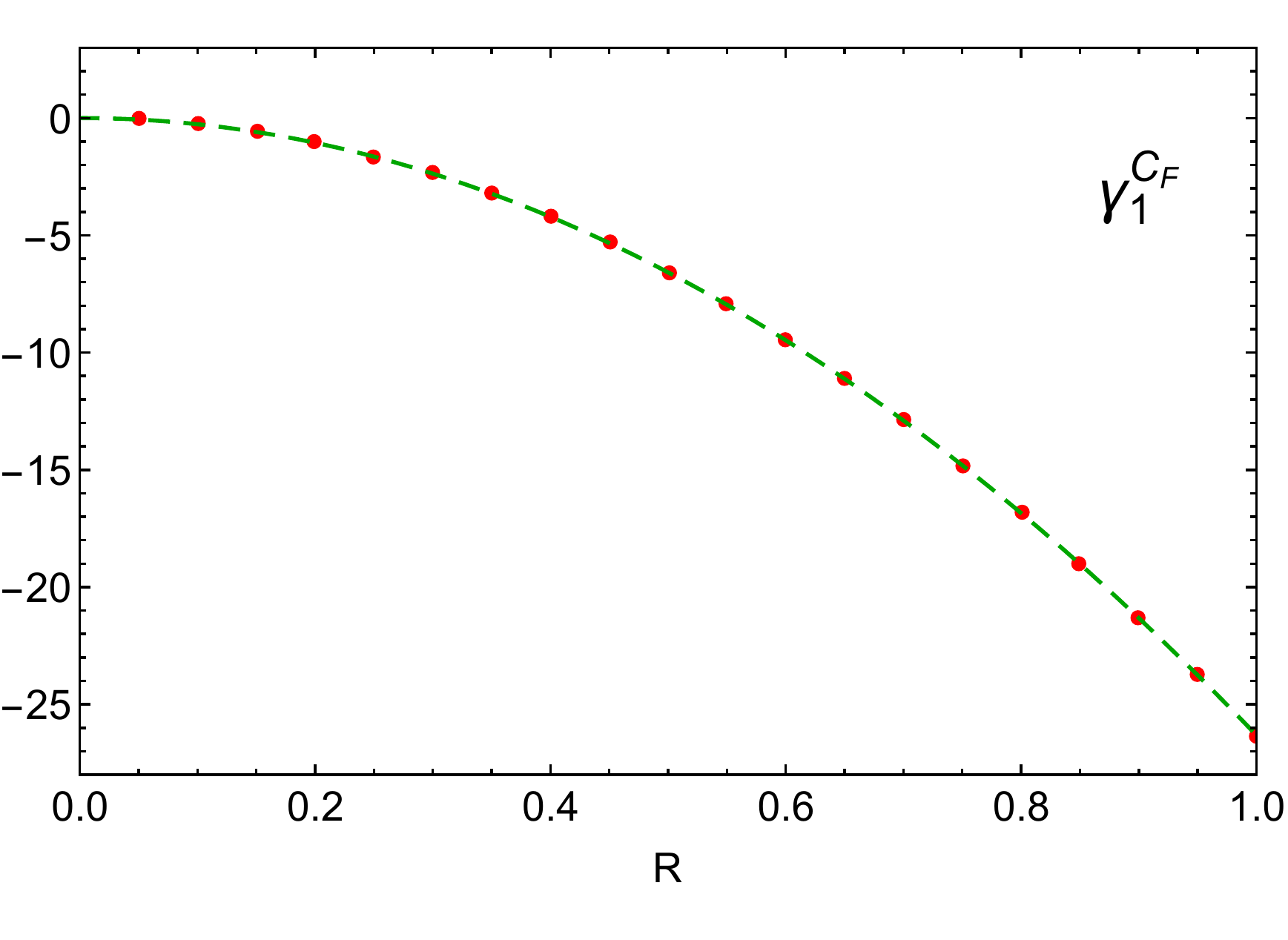}
\includegraphics[width=0.32\textwidth]{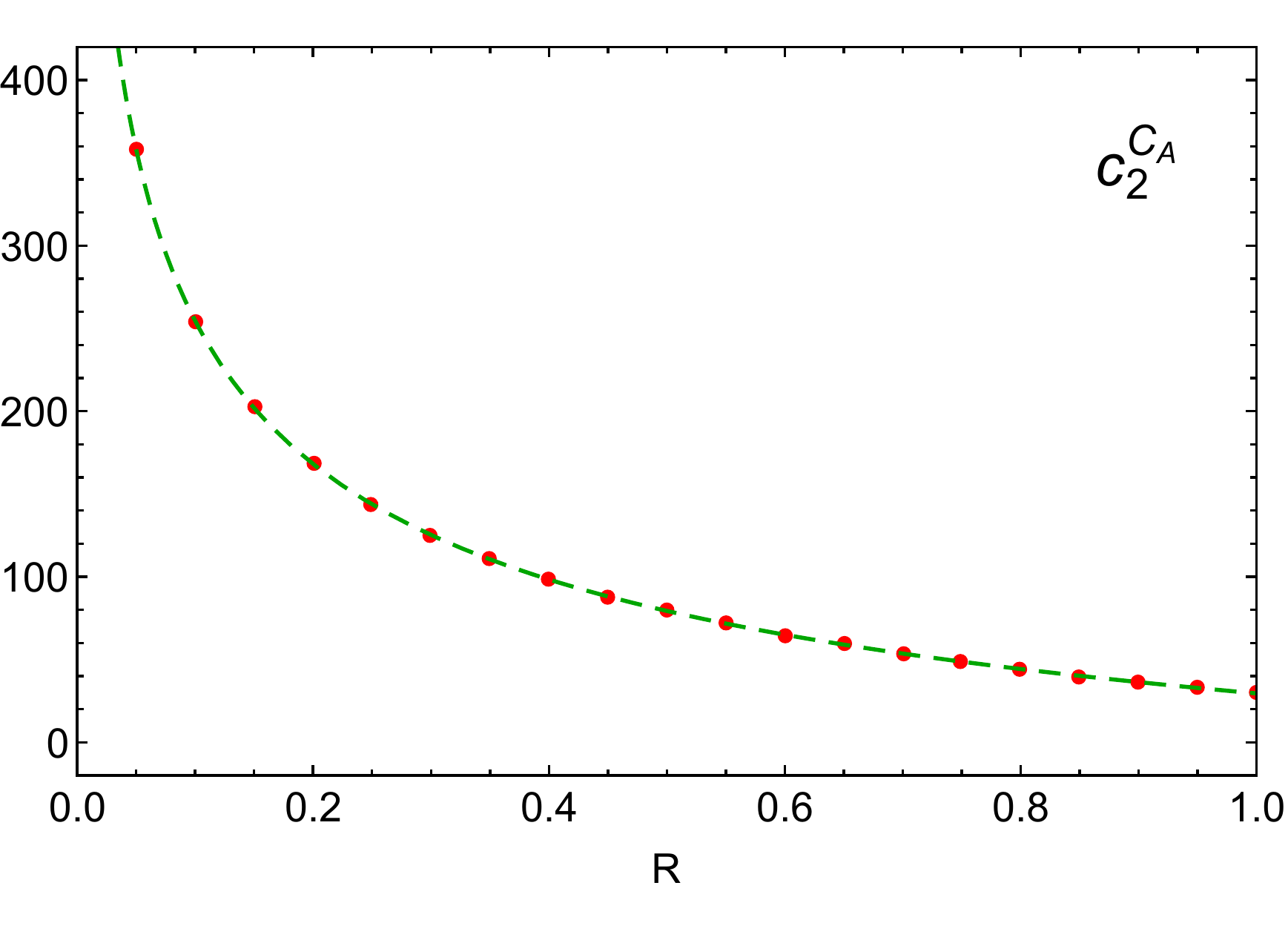}
\includegraphics[width=0.32\textwidth]{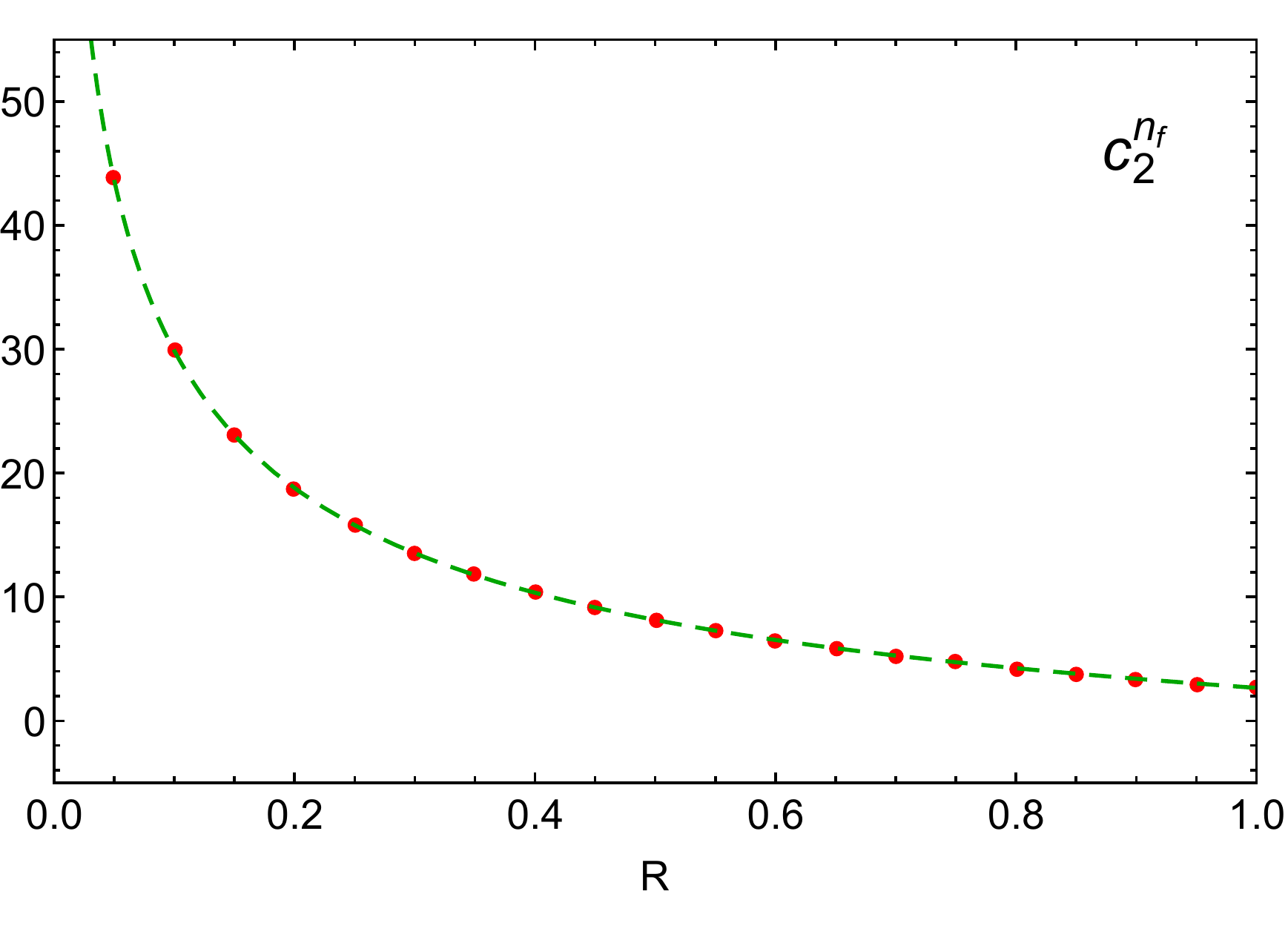}
\includegraphics[width=0.32\textwidth]{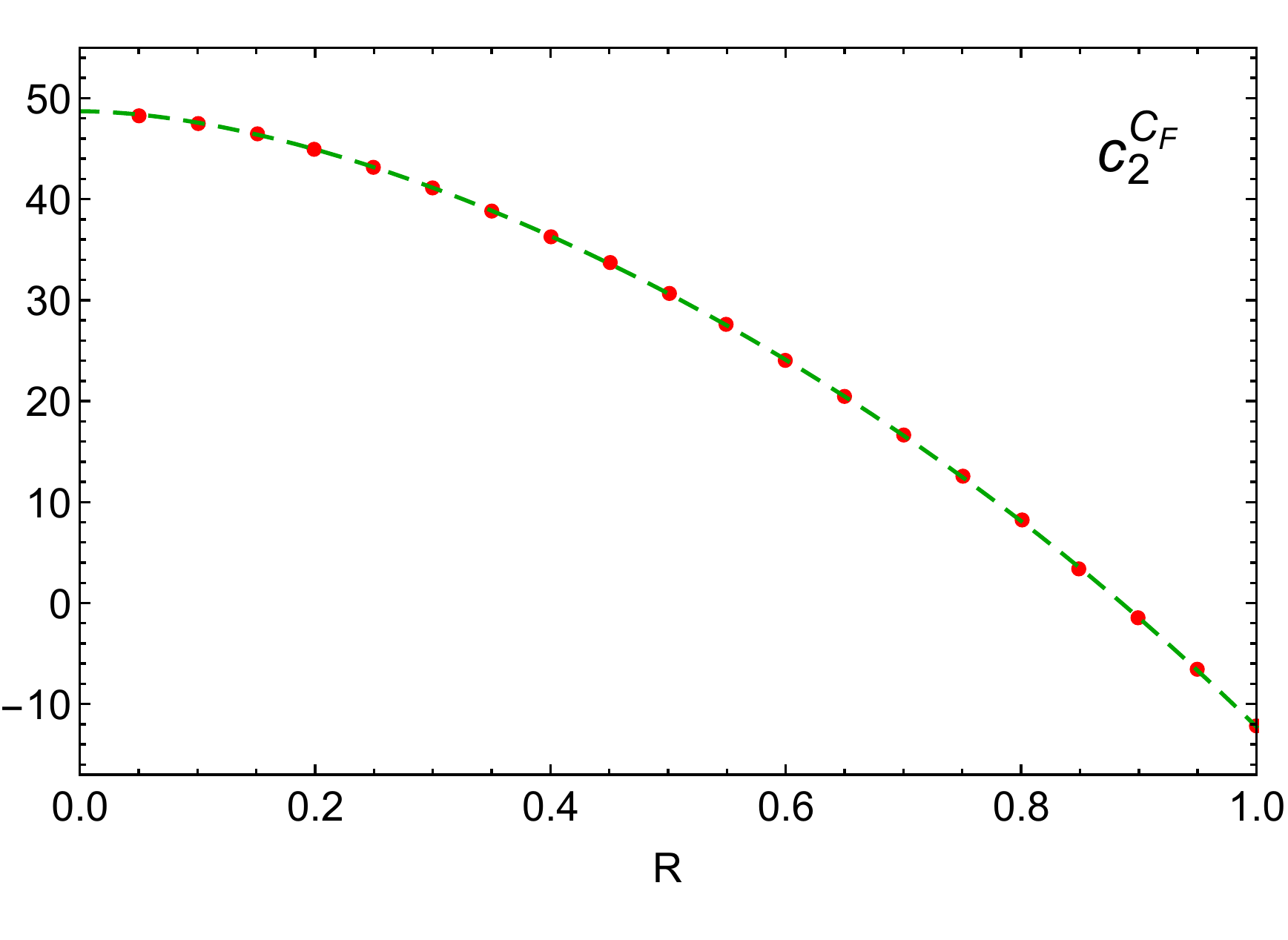}
\caption{Two-loop anomalous dimension and finite term of the renormalised C-parameter jet-veto 
soft function. Red dots indicate values calculated with \softserve~and green dashed lines represent 
the interpolating functions from~\cite{Gangal:2016kuo}.}
\label{fig:CPV}
\end{figure}

The first family of NAE-violating observables are the rapidity-dependent jet vetoes 
from~\cite{Gangal:2014qda}. Specifically, we consider the beam-thrust and C-parameter-like 
jet-veto variables  $\mathcal{T}_{B\rm cm}$ and $\mathcal{T}_{C\rm cm}$ defined in that paper, 
which are both SCET-1 observables with $n=1$. For the C-parameter jet veto, one further has 
$f(y_k,t_k)= 1/(1+y_k)$ and
\begin{align}
F(a,b,y,t_k,t_l,t_{kl}) &= \theta(\Delta_F - R) \;
\text{max}\bigg(\frac{a b}{a(a+b)+(1+a b)y}~,~
\frac{a}{a+b+a(1+a b)y}\bigg) 
\no\\[0.2em]
&\quad +
\theta(R - \Delta_F)\; \frac{1}{1+y}\,,
\end{align}
where $R$ is the jet radius and $\Delta_F= \sqrt{\ln^2 a+\arccos^2(1-2t_{kl})}$, and the 
corresponding expression for the uncorrelated-emission measurement function was given in 
\eqref{eq:G:CPveto}. The jet-veto observables renormalise multiplicatively in cumulant space, 
and therefore the formalism from Section~\ref{sec:cumulant} applies in this case.  Furthermore, 
as the jet algorithm has no effect on a single emission, the NLO coefficients 
$\gamma_0^{C_F}=0$ and $c_1^{C_F}=\pi^2$ are independent of the jet radius $R$, whereas the NNLO 
coefficients are displayed in the range $0\leq R\leq1$ in Figure~\ref{fig:CPV}. From the plots 
it is evident that our \softserve~numbers agree well with the numerical results 
from~\cite{Gangal:2016kuo} indicated by the dashed lines.

\begin{figure}[t]
\centering
\includegraphics[width=0.32\textwidth]{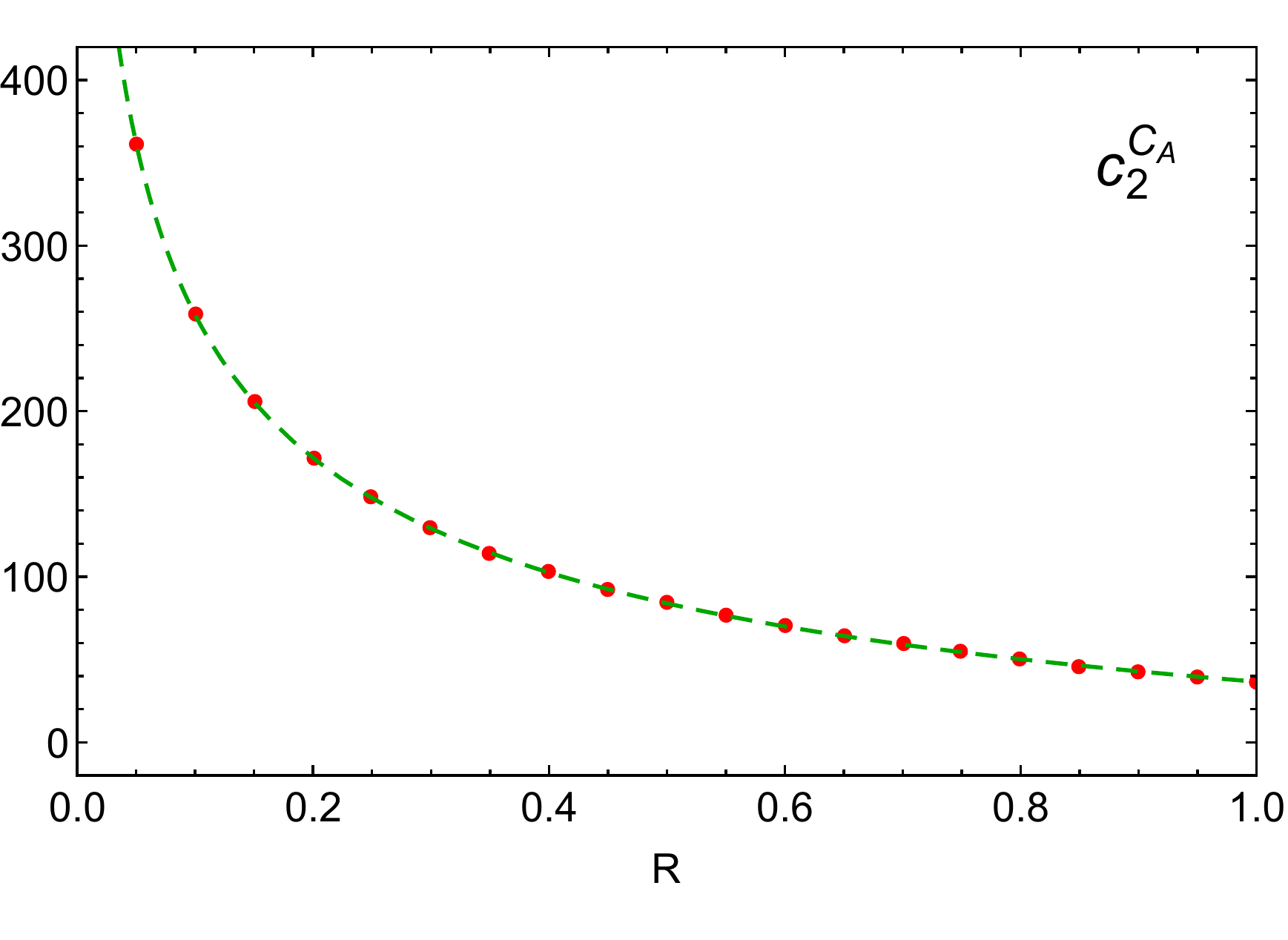}
\includegraphics[width=0.32\textwidth]{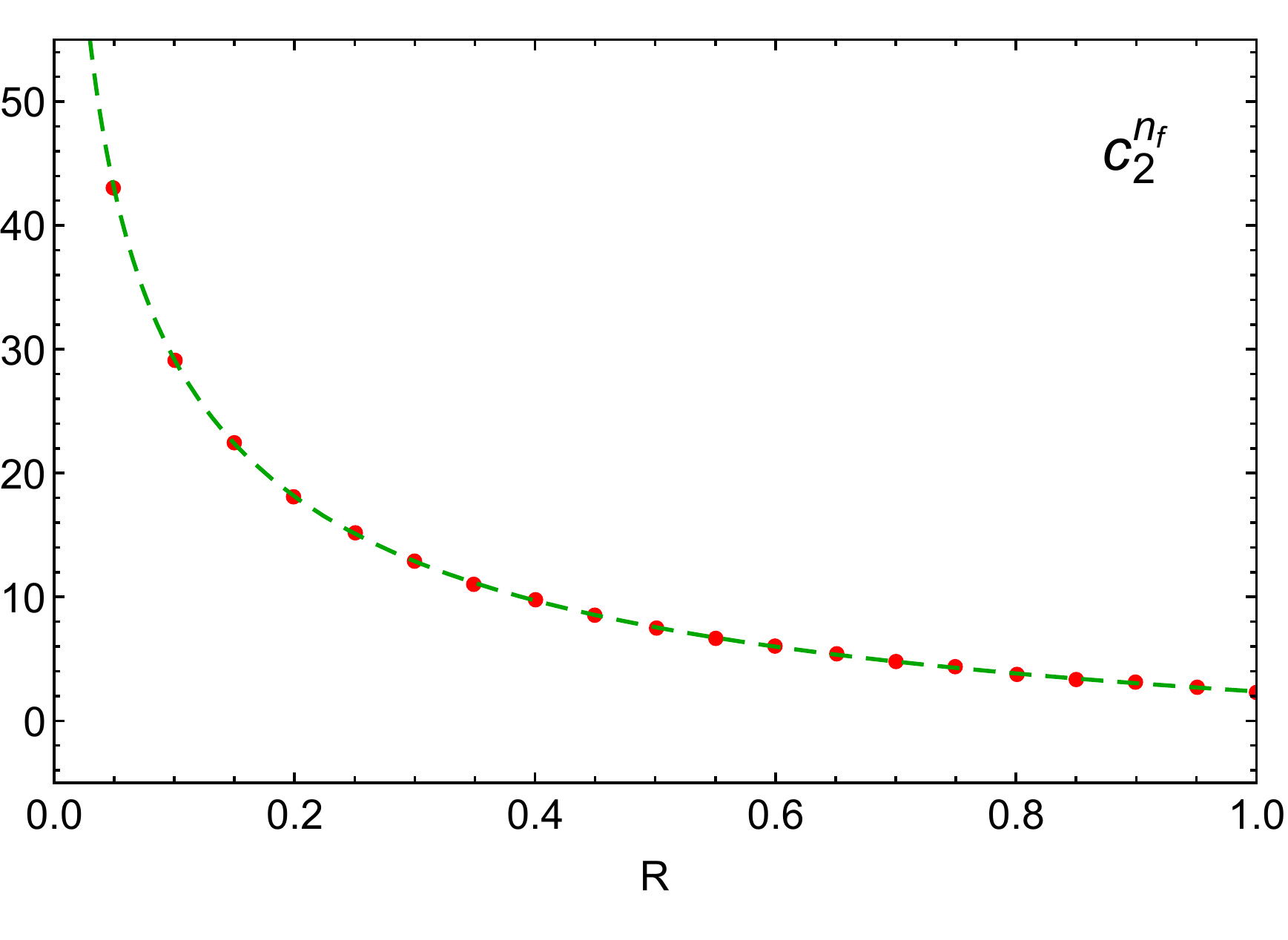}
\includegraphics[width=0.32\textwidth]{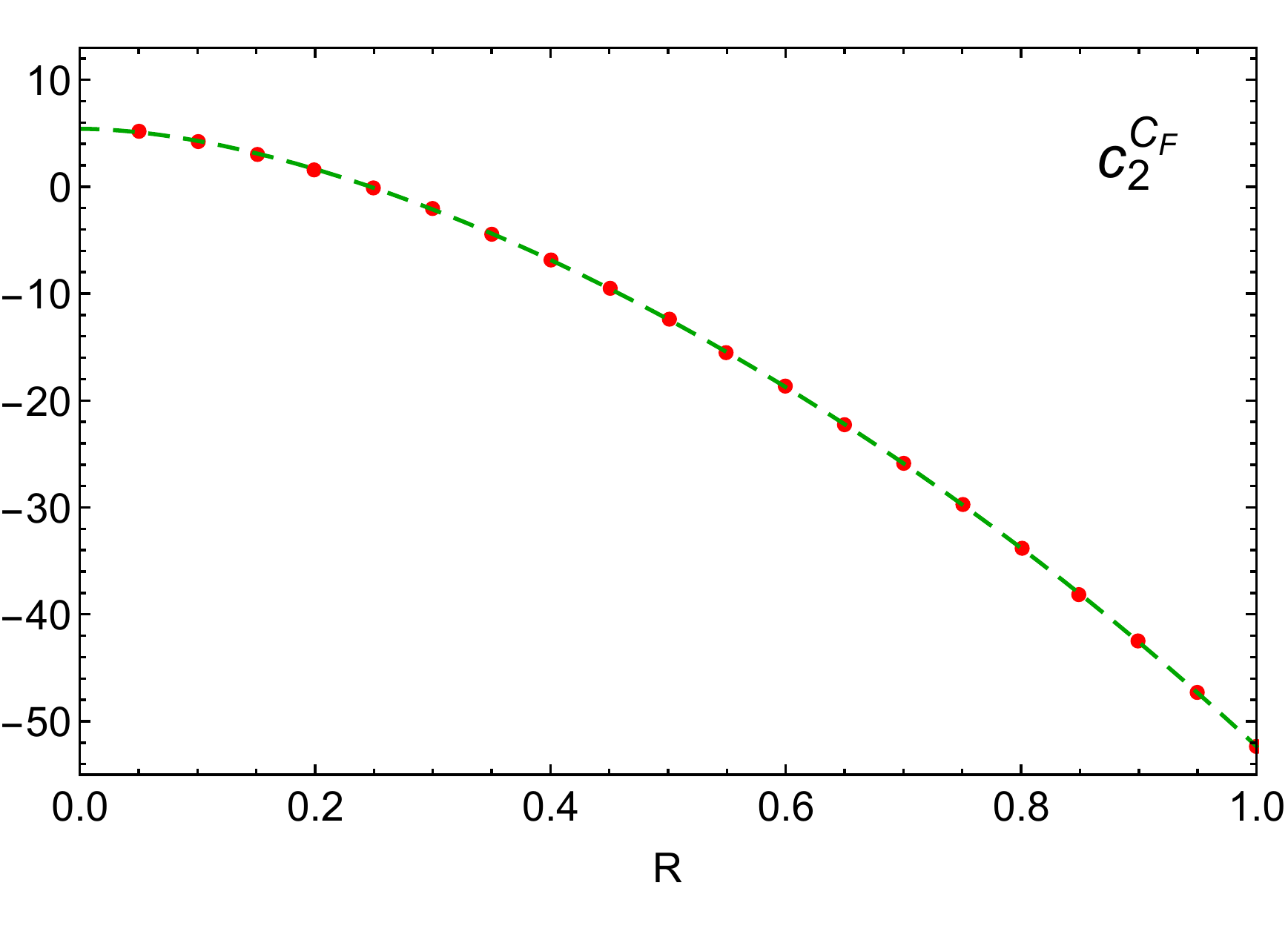}
\caption{The same as in Figure~\ref{fig:CPV} for the
finite term of the renormalised beam-thrust jet-veto soft function.}
\label{fig:TV}
\end{figure}

For the beam-thrust jet veto, the input functions are slightly more complicated and we refer to 
the \softserve~manual for their explicit expressions. As the two jet vetoes have the same 
anomalous dimension, we refrain from showing the corresponding plots in this case, since 
they are -- in view of the negligible numerical uncertainties -- literally identical to the 
upper plots in Figure~\ref{fig:CPV}. The one-loop matching coefficient is, moreover, now given 
by $c_1^{C_F}=\pi^2/3$, and the two-loop coefficients are displayed as a function of the jet 
radius in Figure~\ref{fig:TV}. Our numbers are once more in perfect agreement with the results 
from~\cite{Gangal:2016kuo}. 


\subsubsection*{Standard jet veto}

\begin{figure}[t]
\centering
\includegraphics[width=0.32\textwidth]{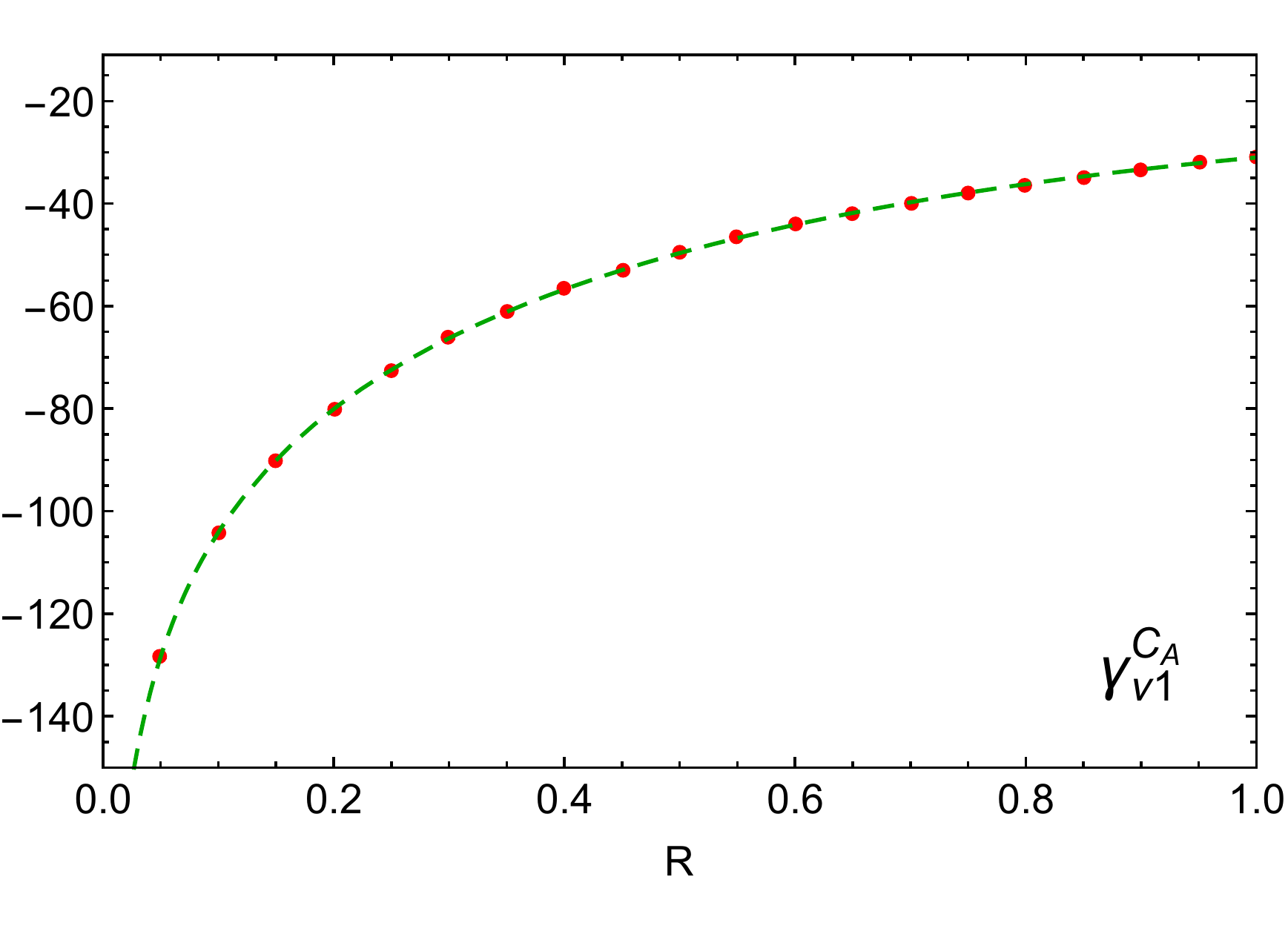}
\includegraphics[width=0.32\textwidth]{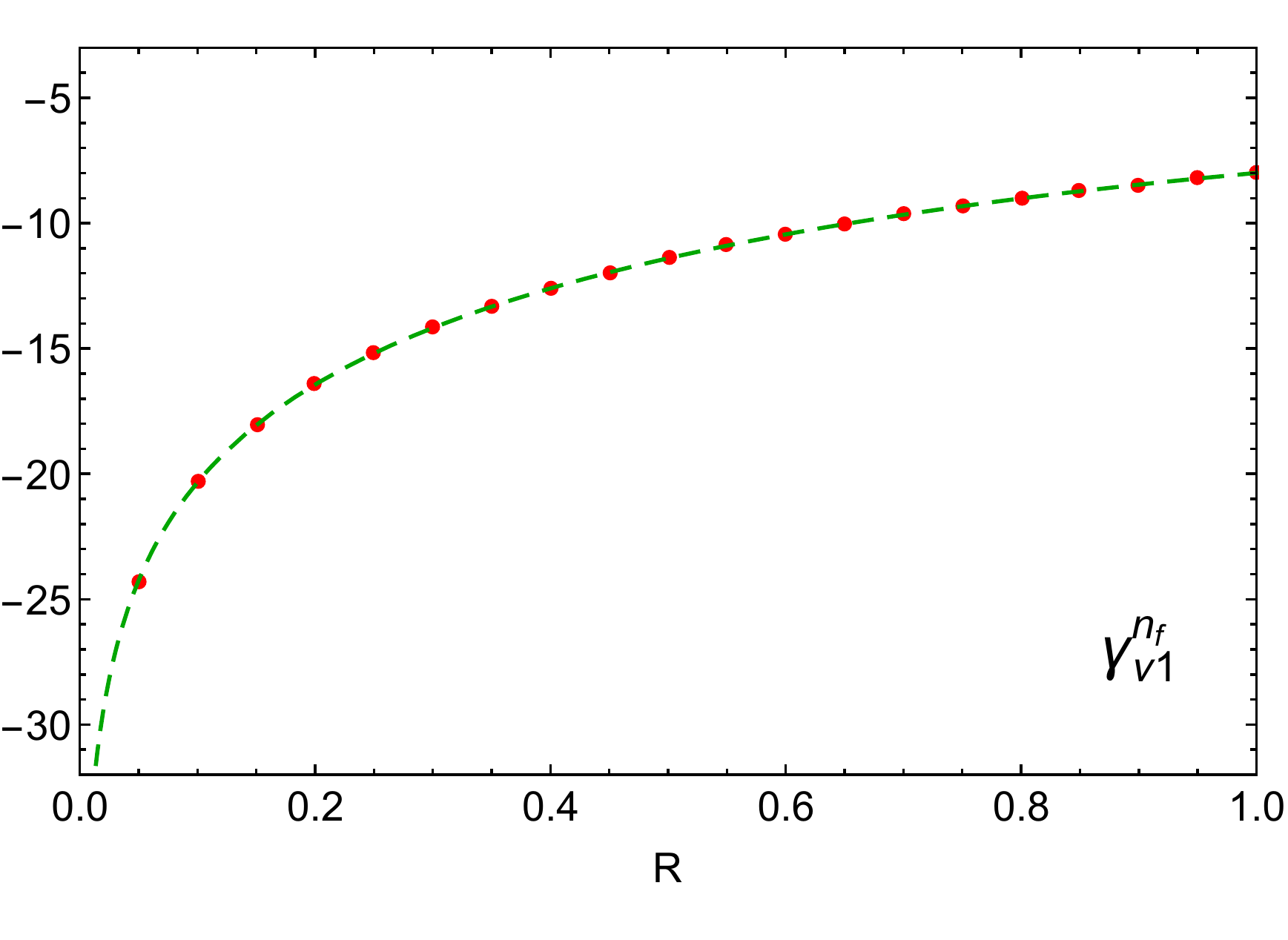}
\includegraphics[width=0.32\textwidth]{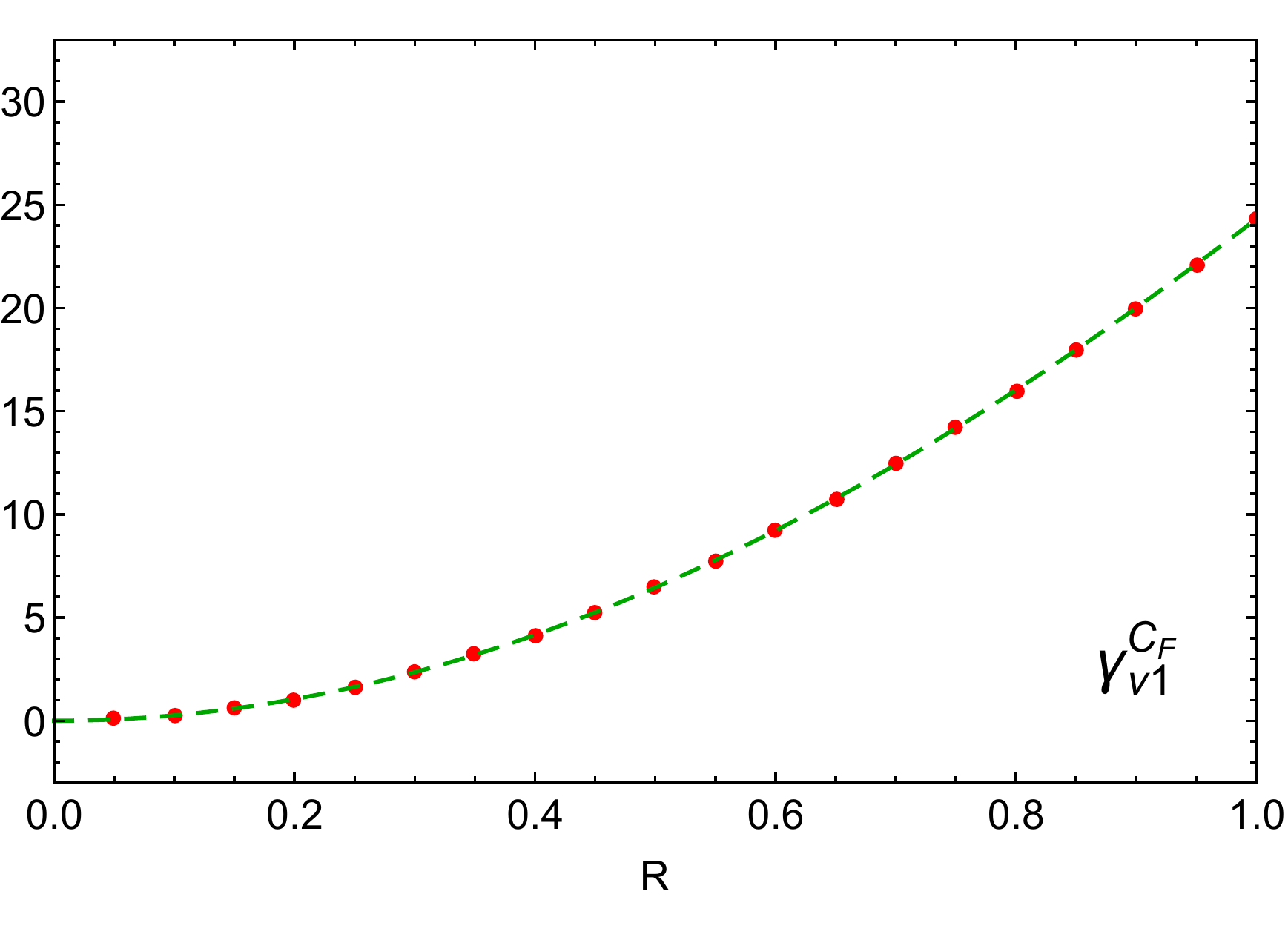}
\includegraphics[width=0.32\textwidth]{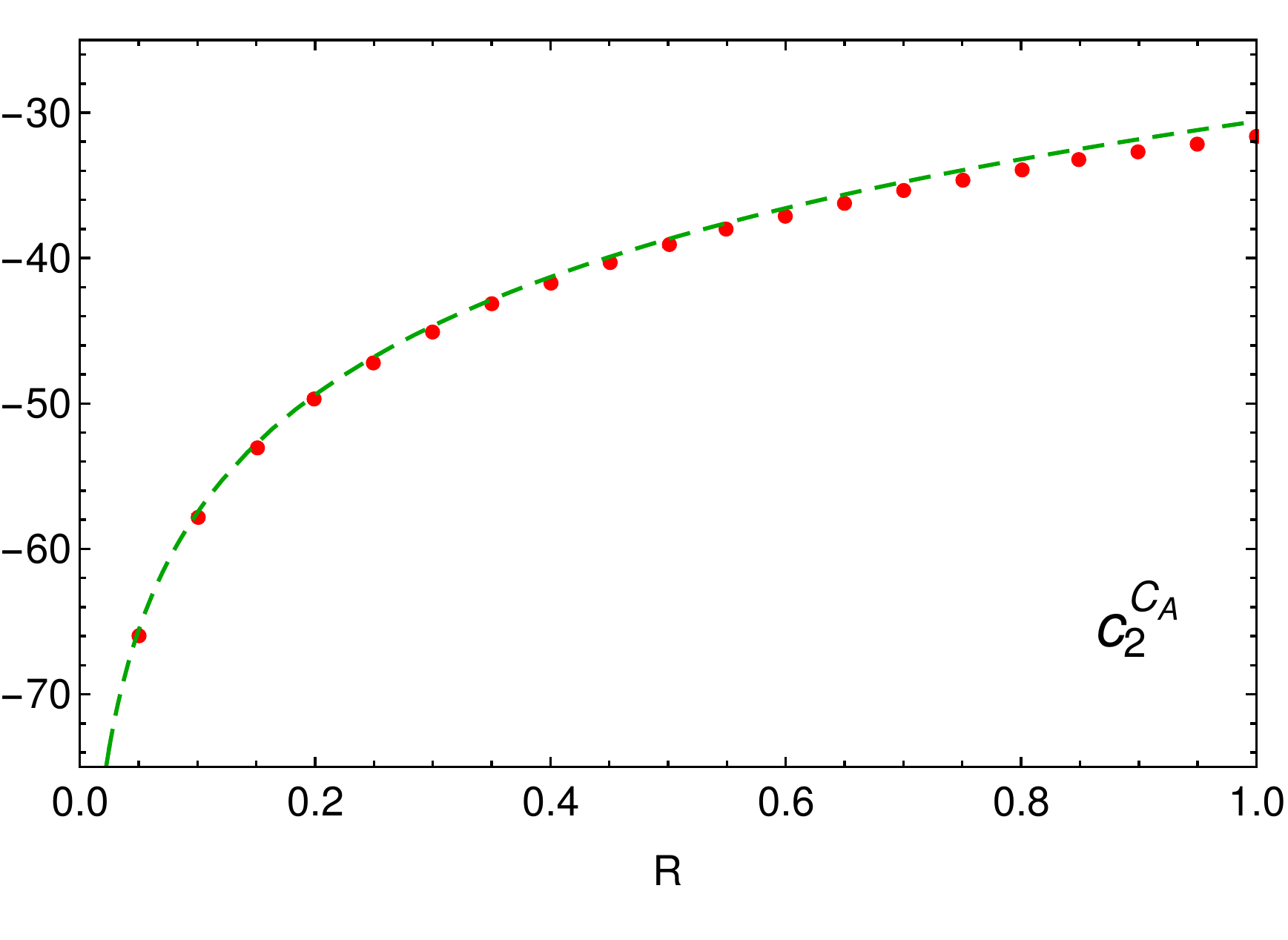}
\includegraphics[width=0.32\textwidth]{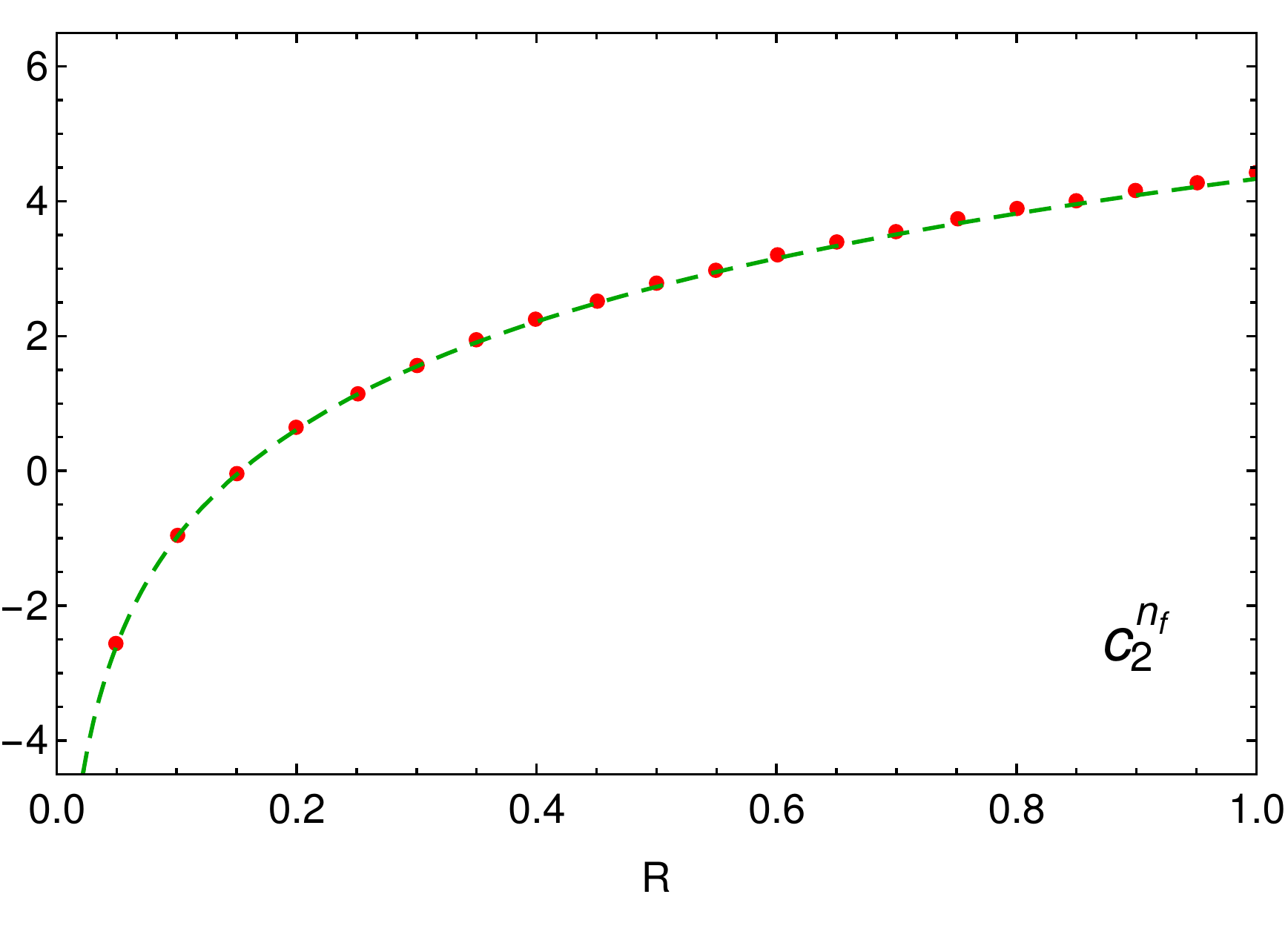}
\includegraphics[width=0.32\textwidth]{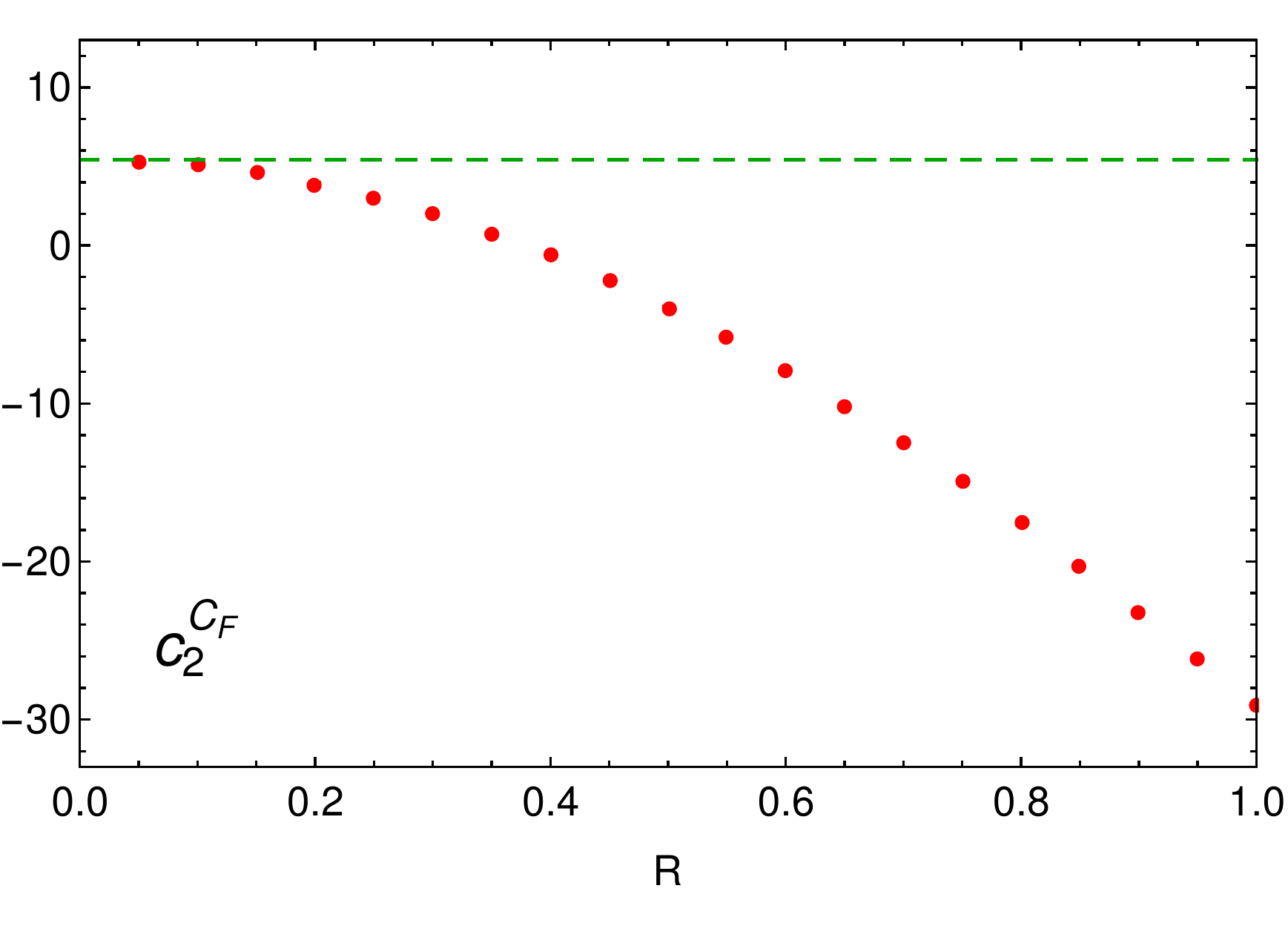}
\caption{Two-loop rapidity anomalous dimension and finite term of the RRG renormalised $p_T$ 
veto soft function. Red dots indicate values calculated with \softserve~and green dashed lines 
show the interpolating functions from~\cite{Becher:2013xia} (upper plots) 
and~\cite{Stewart:2013faa} (lower plots).}
\label{fig:JV}
\end{figure}

The standard way of implementing a jet veto uses a cutoff on the transverse momenta of the 
emissions. The corresponding soft function is in this case defined in SCET-2, and the 
required \softserve~input is given by $n=0$, $f(y_k,t_k)= 1$ and
\begin{align}
F(a,b,y,t_k,t_l,t_{kl}) &= 
\sqrt{\frac{a}{(1+a b)(a+b)}} 
\bigg\{\!
\theta(\Delta_F - R)  
+ \theta(R - \Delta_F)
\sqrt{1+b^2+2b(1-2t_{kl})}\bigg\}\,,
\no\\[0.3em]
 \!\!\!\!G(y_k,y_l,b,t_k,t_l,t_{kl}) &=
 \theta(\Delta_G-R)\, \frac{\max(1,b)}{1+b} 
+ \theta(R-\Delta_G )\,
\frac{ \sqrt{1+b^2+2b(1-2t_{kl})}}{1+b}\,.
\end{align}
As for the rapidity-dependent jet vetoes, the soft function renormalises multiplicatively in 
cumulant space, and the respective NLO coefficients are now given by $\gamma_{\nu,0}^{C_F}=0$ 
and $c_1^{C_F}=-\pi^2/3$. Our numbers for the two-loop rapidity anomalous dimension are shown 
in the upper plots of Figure~\ref{fig:JV}, and they confirm the existing results 
from~\cite{Banfi:2012yh,Becher:2013xia,Stewart:2013faa} indicated by the dashed lines. In the 
RRG setup the two-loop matching corrections can furthermore be compared to~\cite{Stewart:2013faa}, 
which gives these numbers in an expansion in $R\ll1$ up to terms of $\mathcal{O}(R^0)$. As is 
evident from the lower plots in Figure~\ref{fig:JV}, this expansion works surprisingly well 
for the $c_2^{C_A}$ and $c_2^{n_f}$ coefficients even for large values $R\simeq 1$, but it 
misses the leading $\mathcal{O}(R^2)$ correction to $c_2^{C_F}$.

\subsubsection*{Soft-drop jet groomer}

\begin{figure}[htbp]
\centering
\includegraphics[width=0.32\textwidth]{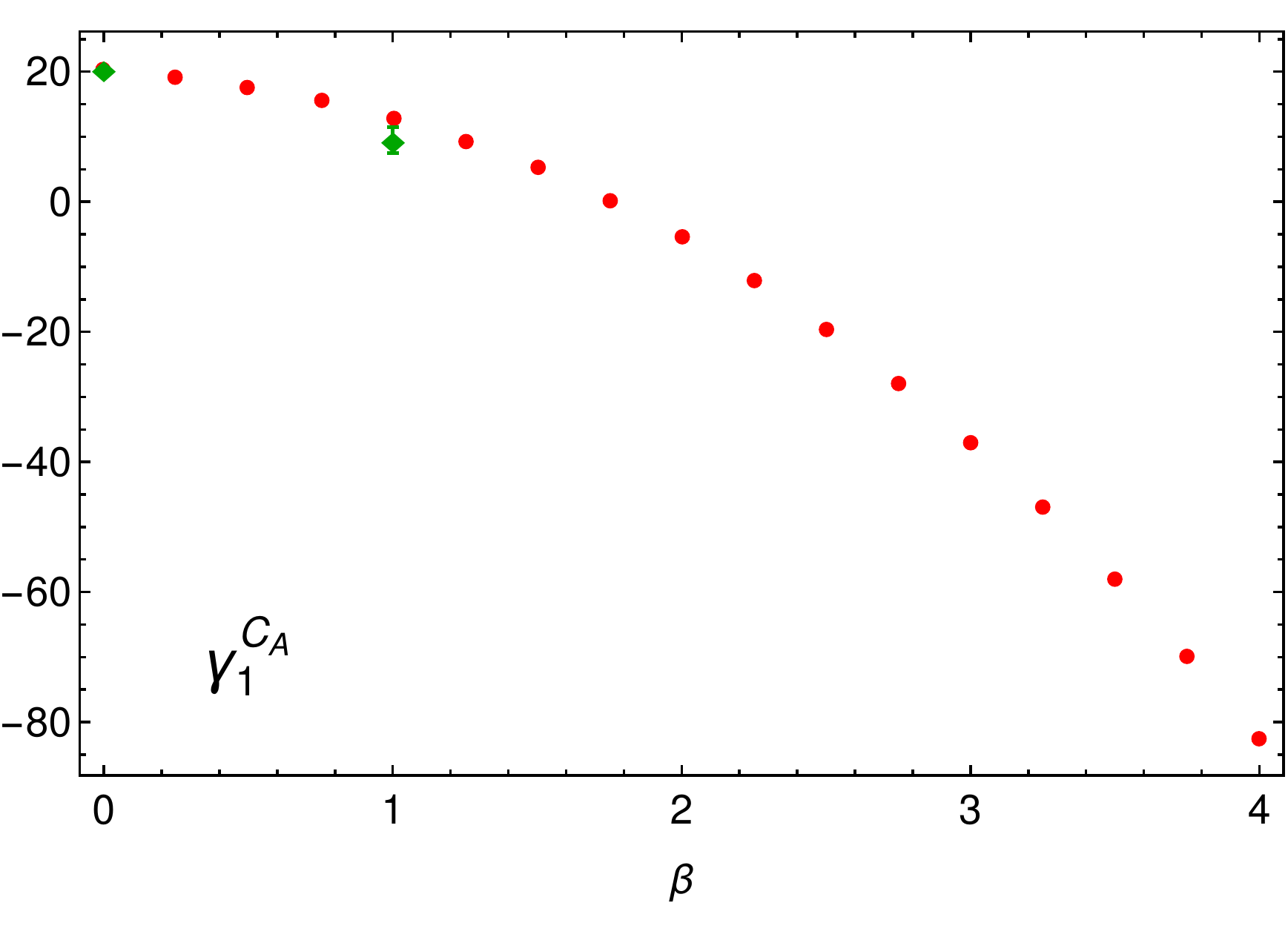}
\includegraphics[width=0.32\textwidth]{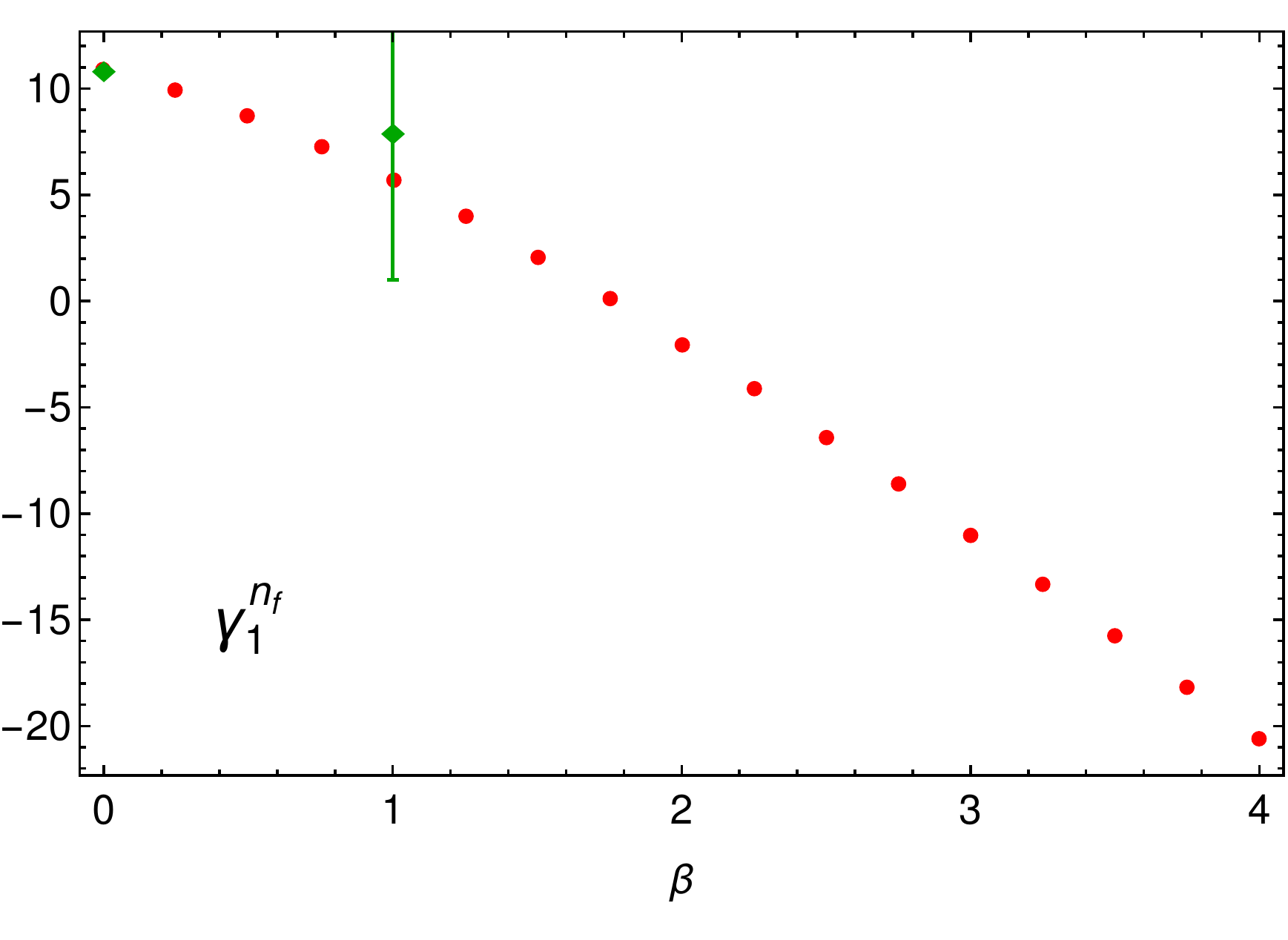}
\includegraphics[width=0.32\textwidth]{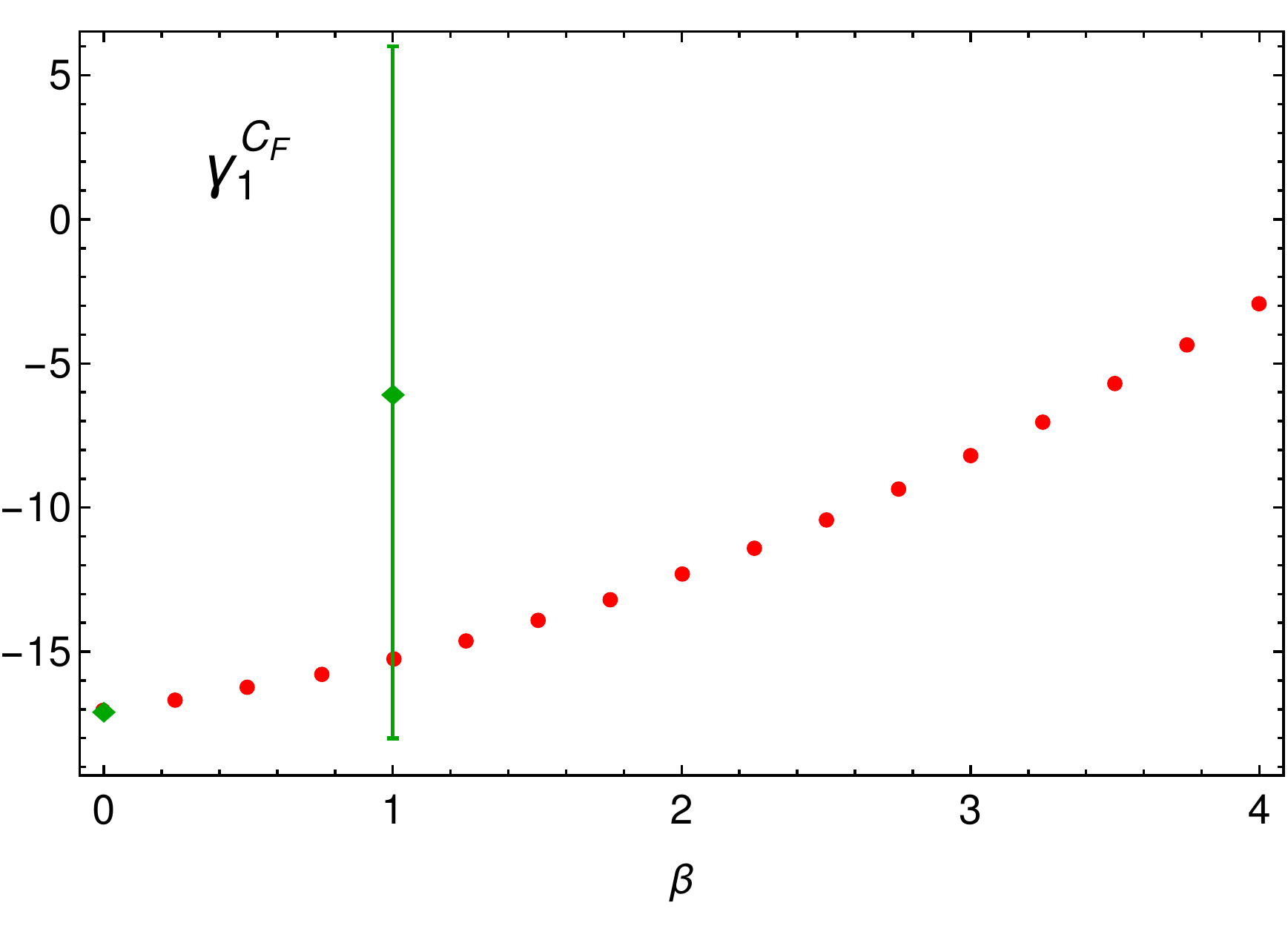}
\includegraphics[width=0.32\textwidth]{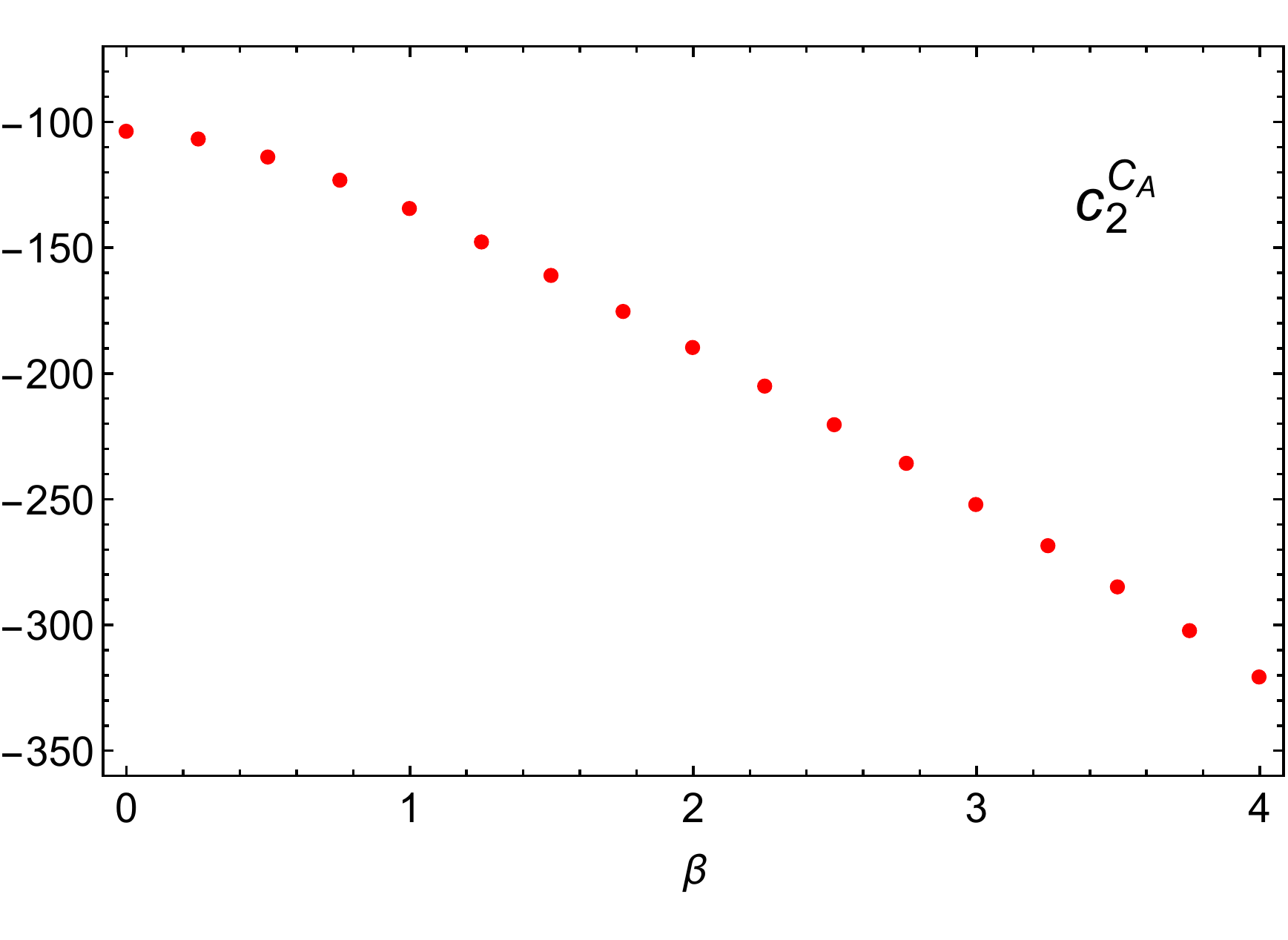}
\includegraphics[width=0.32\textwidth]{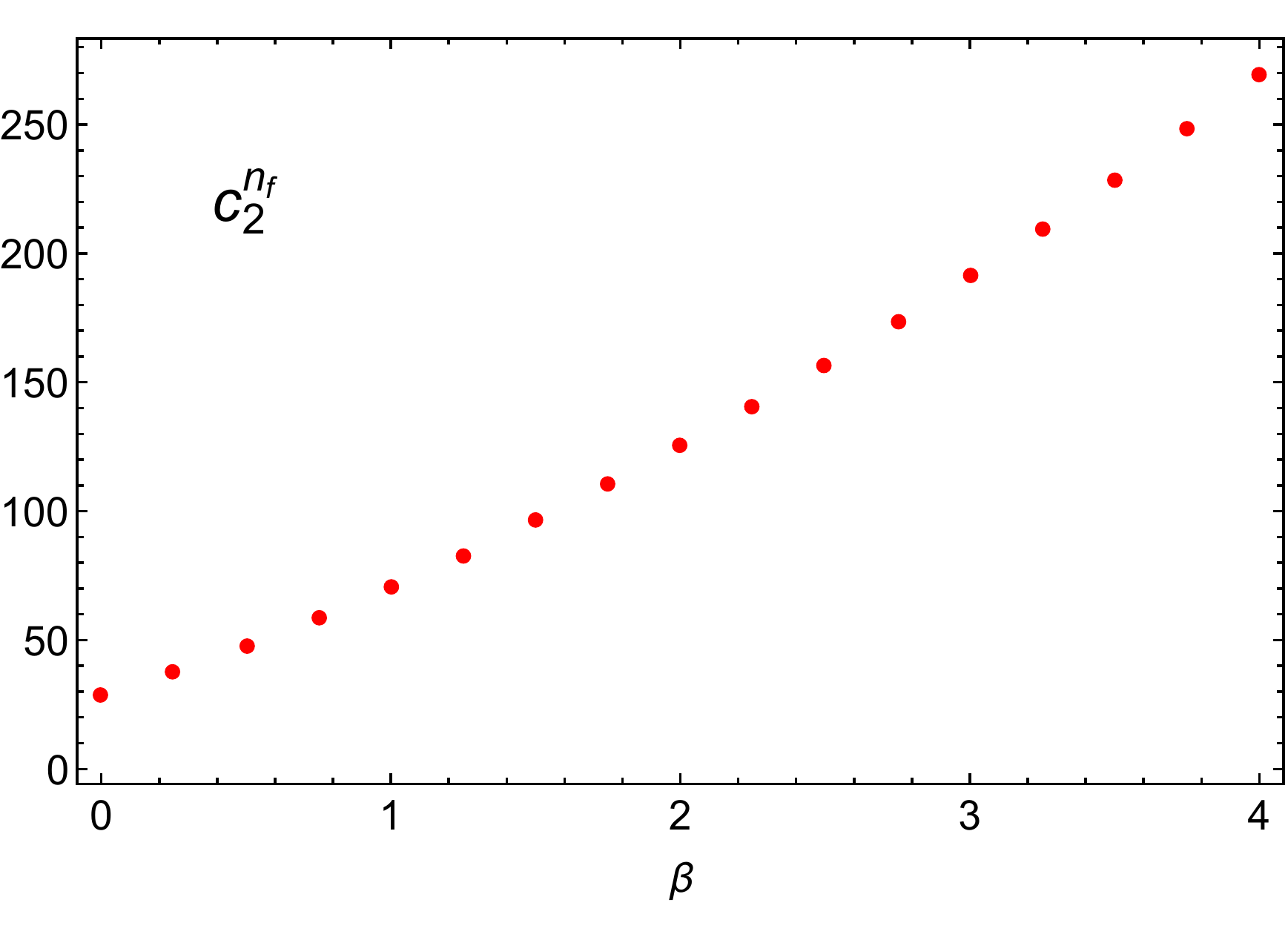}
\includegraphics[width=0.32\textwidth]{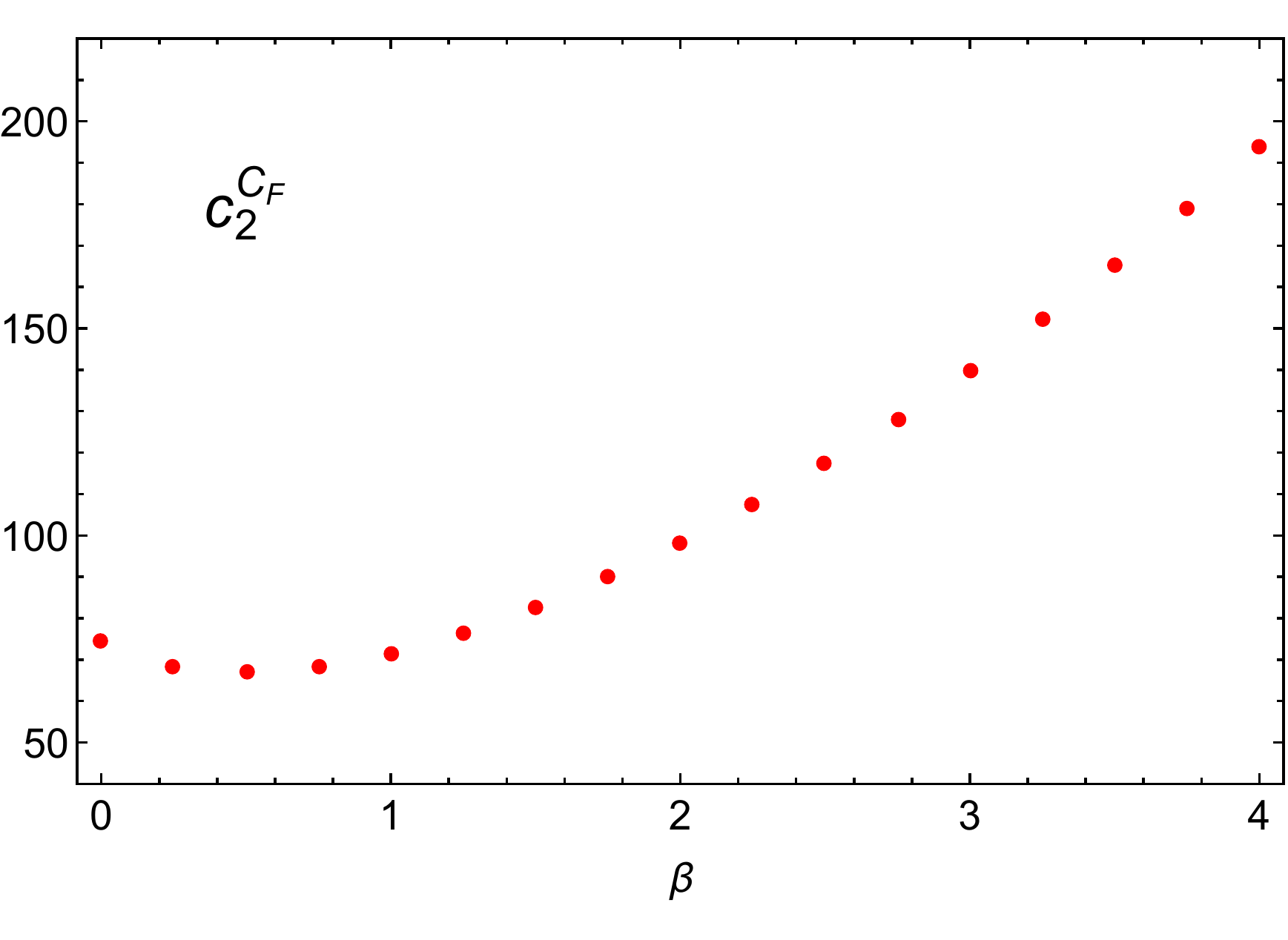}
\caption{Two-loop anomalous dimensions and finite term of the renormalised soft function for 
the soft-drop jet groomer. Red dots indicate values calculated with \softserve~and the green 
diamonds show the numbers from~\cite{Frye:2016aiz}.}
\label{fig:sd}
\end{figure}

Finally, we present novel results for the soft-drop groomed jet mass discussed 
in~\cite{Frye:2016aiz}. According to this definition, the groomer depends on a parameter 
$\beta$, and for values $\beta>0$ considered here, the soft function is defined in SCET-1 
with $n=-1-\beta$. As the formulae for the measurement functions are rather lengthy, we refer to 
the \softserve~distribution for their explicit expressions. The renormalisation of the soft 
function is, moreover, again performed in cumulant space, and the one-loop coefficients are 
found to be $\gamma_0^{C_F}=0$ and $c_1^{C_F}=-\pi^2(3 + 3 \beta + \beta^2)/3 /(1 + \beta)$. 
Our results for the two-loop coefficients are shown in Figure~\ref{fig:sd} together with the 
numbers from~\cite{Frye:2016aiz} for the anomalous dimension. For $\beta=0$ these values have 
been extracted from an analytic calculation, whereas the $\beta=1$ numbers stem from a fit to 
the {\tt EVENT2} generator. From the plots we see that our results confirm these numbers, but 
they are far more precise than the {\tt EVENT2} extraction. Our results for other values of the 
grooming parameter $\beta$ are new, as are the finite terms of the renormalised soft function 
which are shown in the lower plots of the figure.\footnote{As in~\cite{Bell:2018oqa} we validated 
these predictions with independent {\tt pySecDec} runs~\cite{Borowka:2017idc}.}  Our numbers have 
actually already been used to extend the resummation for the soft-drop groomed jet mass to 
next-to-next-to-next-to-leading logarithmic (N$^3$LL) accuracy~\cite{Kardos:2020gty,Kardos:2020ppl}.

\section{Conclusions}
\label{sec:conclude}

We have extended our automated approach for calculating NNLO dijet soft functions to the 
uncorrelated-emission ($C_F^2$) contribution. While one can trivially obtain this term from the 
NLO calculation for observables that obey the NAE theorem, one must calculate it explicitly for 
NAE-violating observables like those that depend on a jet algorithm. From the technical point of 
view, the divergence structure of the $C_F^2$ matrix element differs from the other colour 
structures treated in~\cite{Bell:2018oqa}, and we have devised a novel phase-space parametrisation 
that isolates these singularities.

Our algorithm permits a systematic numerical evaluation of NNLO dijet soft functions, and it is 
implemented in \softserveone~which we release alongside of this paper at 
\url{https://softserve.hepforge.org/}. In addition to the new core routine for calculating the 
uncorrelated-emission contribution to bare dijet soft functions, \softserveone~includes novel 
renormalisation scripts that are compatible with the RRG formalism and observables that 
renormalise directly in momentum space rather than Laplace space.

\softserve~has therefore become a powerful program for calculating NNLO dijet soft functions, 
and we have used it to cross-check existing calculations for multiple $e^+ e^-$ and 
hadron-collider observables, as well as to obtain some novel predictions. In particular, our 
results for the angularity event shape derived in~\cite{Bell:2018oqa} enabled 
NNLL~\cite{Procura:2018zpn} and NNLL$'$~\cite{Bell:2018gce} resummations, and our novel 
predictions for the soft-drop groomed jet mass have recently been employed in a precision 
N$^3$LL resummation in~\cite{Kardos:2020gty,Kardos:2020ppl}. While we hope that \softserve~will 
prove useful for many further applications, an extension of our algorithm to soft functions 
that depend on more than two light-like directions is currently in progress~\cite{Bell:2018mkk}.

\acknowledgments
We thank Andrew Larkoski and Frank Tackmann for discussions.
G.B.~is supported by the Deutsche Forschungsgemeinschaft (DFG) within Research Unit FOR 1873. 
J.T.~acknowledges support from the Villum Fund, project number 00010102, prior funding from DESY 
Hamburg, and also thanks the Albert Einstein Center in Bern for its support and hospitality. 
R.R.~is supported by ERC grant ERC-STG-2015-677323, and acknowledges prior funding from the 
Swiss National Science Foundation (SNF) under grants CRSII2$\_$160814 and 200020$\_$182038.

\appendix

\section{Anomalous dimensions}
\label{app:AD}

In this appendix we rederive the integral representations from~\cite{Bell:2018vaa}, which allow 
for a fast evaluation of the soft anomalous dimension $\gamma^{S}$ and the collinear anomaly 
exponent $\mathcal{F}(\tau,\mu)$ for SCET-1 and SCET-2 observables, respectively. While our 
derivation follows the conventions from~\cite{Bell:2018vaa}, it differs in one aspect from that 
work; namely the sector decomposition step in \eqref{eq:rr:ga12} is performed only for the 
same-hemisphere contribution for uncorrelated emissions (region A), while it was also applied 
to the opposite-hemisphere case (region B) in~\cite{Bell:2018vaa}.

We start with the $C_F^2$ contribution to the soft anomalous dimension for SCET-1 observables, 
for which (22) of~\cite{Bell:2018vaa} is replaced by
\begin{align}
\gamma^{C_F}_1 &=
 \frac{128}{\pi}\,
\int_0^1 \!dy \int_0^1 \!dt_l \;\,
\frac{1}{\altsqrt{4t_l\bar t_l}}\;\,\frac{1}{y}\;
\ln^2 \left( \frac{(1+y)^n\, f(y,t_l)}{f(0,t_l)} \right) 
\nonumber\\[0.2em]
&\quad
+\frac{256}{\pi}\,
\int_0^1 \!dy \int_0^1 \!dt_l \;\,
\frac{\ln f(0,t_l)}{\altsqrt{4t_l\bar t_l}}\;
 \frac{\ln f(y,t_l)}{y_+} 
\nonumber\\[0.2em]
&\quad
-\frac{512}{\pi^2}\,
\int_0^1 \!dt_k \;\,
\frac{\ln f(0,t_k)}{\altsqrt{4t_k\bar t_k}}\;
\int_0^1 \!dy \int_0^1 \!dt_l \;\,
\frac{1}{\altsqrt{4t_l\bar t_l}} \; 
\frac{\ln f(y,t_l)}{y_+}
\nonumber\\[0.2em]
&\quad
-\frac{128}{\pi^2}
\int_0^1 \!dy \int_0^1 \!db \int_0^1 \!dt_l \int_0^1 \!dt_{kl} \;
\frac{1}{\altsqrt{16t_l\bar t_l t_{kl}\bar t_{kl}}}\; \,
\frac{\mathcal{H}_1(y,b,t_l,t_{kl})}{y_+b_+}\,
\nonumber\\[0.3em]
&\quad
-\frac{64}{\pi^2}
\int_0^1 \!dr \int_0^1 \!db \int_0^1 \!dt_l \int_0^1 \!dt_{kl} \;
\frac{1}{\altsqrt{16t_l\bar t_l t_{kl}\bar t_{kl}}}\; \,
\frac{\mathcal{H}_2(r,b,t_l,t_{kl})}{r_+b_+}
\nonumber\\[0.2em]
&\quad
-\frac{128}{\pi^2}
\int_0^1 \!dy_k \int_0^1 \!db \int_0^1 \!dt_l \int_0^1 \!dt_{kl} \;
\frac{1}{\altsqrt{16t_l\bar t_l t_{kl}\bar t_{kl}}}\; \,
\frac{\mathcal{H}_3(y_k,b,t_l,t_{kl})}{{y_k}_+b_+}\,
\nonumber\\[0.3em]
&\quad
-\frac{128}{\pi^2}
\int_0^1 \!dy_l \int_0^1 \!db \int_0^1 \!dt_l \int_0^1 \!dt_{kl} \;
\frac{1}{\altsqrt{16t_l\bar t_l t_{kl}\bar t_{kl}}}\; \,
\frac{\mathcal{H}_4(y_l,b,t_l,t_{kl})}{{y_l}_+b_+}\,,
\label{eq:CF2:gamma1}
\end{align}
with
\begin{align}
\mathcal{H}_{1}(y,b,t_l,t_{kl}) &=
\ln G_{A_1}(y,0,b,t_k^+,t_l,t_{kl})
+ \ln G_{A_2}(y,0,b,t_k^+,t_l,t_{kl})
+ (t_k^+\to t_k^-)\,,
\nonumber\\
\mathcal{H}_{2}(r,b,t_l,t_{kl}) &=
\ln G_{A_1}(0,r,b,t_k^+,t_l,t_{kl})
+ \ln G_{A_2}(0,r,b,t_k^+,t_l,t_{kl})
+ (t_k^+\to t_k^-)\,,
\nonumber\\
\mathcal{H}_{3}(y_k,b,t_l,t_{kl}) &=
\ln G_{B}(y_k,0,b,t_k^+,t_l,t_{kl})
+ (t_k^+\to t_k^-)\,,
\nonumber\\
\mathcal{H}_{4}(y_l,b,t_l,t_{kl}) &=
\ln G_{B}(0,y_l,b,t_k^+,t_l,t_{kl})
+ (t_k^+\to t_k^-)\,,
\end{align} 
and
\begin{align}
t_k^\pm = t_l + t_{kl} - 2 t_l t_{kl} \pm 2 \sqrt{t_l\bar t_l t_{kl}\bar t_{kl}}\,.
\end{align}
Similar to~\cite{Bell:2018vaa}, we find that this result only holds if the following constraint 
\begin{align}
& 
\frac{8}{\pi}\,
\int_0^1 \!dt_l \;\,
\frac{\ln^2 f(0,t_l)}{\altsqrt{4t_l\bar t_l}}
-\frac{16}{\pi^2}\,
\int_0^1 \!dt_k \;\,
\frac{\ln f(0,t_k)}{\altsqrt{4t_k\bar t_k}}\;
\int_0^1 \!dt_l \;\,
\frac{\ln f(0,t_l)}{\altsqrt{4t_l\bar t_l}}
\nonumber\\[0.2em]
&\qquad
-\frac{4}{\pi^2}
\int_0^1 \!db \int_0^1 \!dt_l \int_0^1 \!dt_{kl} \;
\frac{1}{\altsqrt{16t_l\bar t_l t_{kl}\bar t_{kl}}}\; \,
\frac{\mathcal{H}_0(b,t_l,t_{kl})}{b_+}
\,=\,0
\label{eq:CF2:constraint}
\end{align}
is satisfied, where
\begin{align}
\mathcal{H}_{0}(b,t_l,t_{kl}) &=
\ln G_{A_1}(0,0,b,t_k^+,t_l,t_{kl})
+ \ln G_{A_2}(0,0,b,t_k^+,t_l,t_{kl})
\nonumber\\
&\quad
+ 2\ln G_{B}(0,0,b,t_k^+,t_l,t_{kl})
+ (t_k^+\to t_k^-)\,.
\end{align} 
Moreover, we find an additional contribution to the soft anomalous dimension, which we 
conjecture to vanish for all observables,  given by
\begin{align}
\label{eq:CF2:deltagamma1}
\Delta \gamma^{C_F}_1 &=
\frac{64}{n} \,\bigg\{ 
\frac{4}{\pi}\,
\int_0^1 \!dt_l \;\,
\frac{\ln^3 f(0,t_l)}{\altsqrt{4t_l\bar t_l}}
-\frac{2}{\pi}\,
\int_0^1 \!dt_l \;\,
\frac{\ln(16t_l\bar t_l)}{\altsqrt{4t_l\bar t_l}}\;\ln^2 f(0,t_l)
\\[0.2em]
&\qquad\quad
-\frac{8}{\pi^2}\,
\int_0^1 \!dt_k \;\,
\frac{\ln f(0,t_k)}{\altsqrt{4t_k\bar t_k}}\;
\int_0^1 \!dt_l \;\,
\frac{\ln f(0,t_l)}{\altsqrt{4t_l\bar t_l}}\;
\ln \frac{f(0,t_l)}{16t_l\bar t_l}
\nonumber\\[0.2em]
&\qquad\quad
+\frac{1}{\pi^2}
\int_0^1 \!db \int_0^1 \!dt_l \int_0^1 \!dt_{kl} \;
\frac{1}{\altsqrt{16t_l\bar t_l t_{kl}\bar t_{kl}}}\; \,
\bigg[\frac{1}{b} \,\ln \frac{256\,t_l\bar t_l t_{kl}\bar t_{kl}\,b^2}{(1+b)^4} \bigg]_+\;
\mathcal{H}_0(b,t_l,t_{kl})
\nonumber\\[0.2em]
&\qquad\quad
-\frac{2}{\pi^2}
\int_0^1 \!db \int_0^1 \!dt_l \int_0^1 \!dt_{kl} \;
\frac{1}{\altsqrt{16t_l\bar t_l t_{kl}\bar t_{kl}}}\; \,
\frac{\mathcal{H}_5(b,t_l,t_{kl})}{b_+}
\nonumber\\[0.2em]
&\qquad\quad
+\frac{2}{\pi^2}
\int_0^1 \!db \int_0^1 \!dt_l \int_0^1 \!dt_{kl} \int_0^1 \!ds \;
\frac{1}{\altsqrt{16t_l\bar t_l t_{kl}\bar t_{kl}}}\; \,
\frac{1}{b} \bigg[\frac{1}{s\sqrt{1-s^2}} \bigg]_+ \;\mathcal{H}_6(b,t_l,t_{kl},s)\bigg\},
\nonumber
\end{align}
with
\begin{align}
\mathcal{H}_{5}(b,t_l,t_{kl}) &=
\ln^2 G_{A_1}(0,0,b,t_k^+,t_l,t_{kl})
+ \ln^2 G_{A_2}(0,0,b,t_k^+,t_l,t_{kl})
\nonumber\\
&\quad
+ 2\ln^2 G_{B}(0,0,b,t_k^+,t_l,t_{kl})
+ (t_k^+\to t_k^-)\,,
\nonumber\\
\mathcal{H}_{6}(b,t_l,t_{kl},s) &=
\ln G_{A_1}(0,0,b,t_k^{\oplus},t_l,t_{kl})
+ \ln G_{A_2}(0,0,b,t_k^{\oplus},t_l,t_{kl})
\nonumber\\
&\quad
+ 2\ln G_{B}(0,0,b,t_k^{\oplus},t_l,t_{kl})
+ (t_k^{\oplus}\to t_k^{\ominus})\,,
\end{align} 
and
\begin{align}
t_k^{\oplus} &= t_l + t_{kl} - 2 t_l t_{kl} + 2 \sqrt{t_l\bar t_l t_{kl}\bar t_{kl}(1-s^2)}\,,
\nonumber\\
t_k^{\ominus} &= t_l + t_{kl} - 2 t_l t_{kl} - 2 \sqrt{t_l\bar t_l t_{kl}\bar t_{kl}(1-s^2)}\,.
\end{align}

\noindent
For SCET-2 observables the relevant formulae are $d_2^{\,C_F} = -\gamma^{C_F}_1$,
\begin{align}
\Delta d^{\,C_F}_2 &=
64 \,\bigg\{ 
-\frac{4}{\pi}\,
\int_0^1 \!dt_l \;\,
\frac{\ln^3 f(0,t_l)}{\altsqrt{4t_l\bar t_l}}
+\frac{8}{\pi^2}\,
\int_0^1 \!dt_k \;\,
\frac{\ln f(0,t_k)}{\altsqrt{4t_k\bar t_k}}\;
\int_0^1 \!dt_l \;\,
\frac{\ln^2 f(0,t_l)}{\altsqrt{4t_l\bar t_l}}\;
\nonumber\\[0.2em]
&\qquad\quad
-\frac{1}{\pi^2}
\int_0^1 \!db \int_0^1 \!dt_l \int_0^1 \!dt_{kl} \;
\frac{1}{\altsqrt{16t_l\bar t_l t_{kl}\bar t_{kl}}}\; \,
\bigg[\frac{1}{b} \,\ln \frac{b^2}{(1+b)^4} \bigg]_+\;\mathcal{H}_0(b,t_l,t_{kl})
\nonumber\\[0.2em]
&\qquad\quad
+\frac{2}{\pi^2}
\int_0^1 \!db \int_0^1 \!dt_l \int_0^1 \!dt_{kl} \;
\frac{1}{\altsqrt{16t_l\bar t_l t_{kl}\bar t_{kl}}}\; \,
\frac{\mathcal{H}_5(b,t_l,t_{kl})}{b_+}\bigg\},
\label{eq:CF2:deltad2} 
\end{align}
and the same constraint \eqref{eq:CF2:constraint} has to be fulfilled. According 
to~\cite{Bell:2018vaa}, these relations are slightly modified for cumulant soft functions, and
we will not repeat the required changes here.

While the above formulae hold under the assumptions specified in Section~\ref{sec:measure}, 
our \softserve~implementation is subject to one additional constraint, i.e.~the measurement 
function $\omega(\lbrace k_{i} \rbrace)$ must be strictly real and non-negative. The 
\softserve~routines \texttt{ADLap} and \texttt{ADMom} therefore cannot immediately be applied 
to Fourier-space soft functions, but as we explained in Appendix B of~\cite{Bell:2018oqa}, there 
exists a workaround in \softserve, which consists in replacing the complex-valued measurement 
functions by their absolute values, and by multiplying the result with appropriate factors that 
reshuffle the expansion in the dimensional and rapidity regulators. For the anomalous dimensions 
considered here, there exists a similar workaround, and in the SCET-1 case one finds that the 
anomalous dimension in \eqref{eq:CF2:gamma1} is not changed, whereas \eqref{eq:CF2:constraint} 
and \eqref{eq:CF2:deltagamma1}  receive additional contributions in this case given by $(-\pi^2)$ 
and $-128\pi/n \int_0^1 \!dt_l/\altsqrt{4t_l\bar t_l}\,\ln f(0,t_l)$, respectively. For SCET-2 
soft functions, we find that the collinear anomaly exponent itself is shifted by 
$-2\pi^2\beta_0 C_F$, whereas \eqref{eq:CF2:constraint} and \eqref{eq:CF2:deltad2} are changed 
by $(-\pi^2)$ and $128\pi \int_0^1 \!dt_l/\altsqrt{4t_l\bar t_l}\,\ln f(0,t_l)$. 


\hypersetup{
    urlcolor = .
}

\bibliography{bibliography}

\providecommand{\href}[2]{#2}\begingroup\raggedright\begin{thebibliography}{10}

\bibitem{Bell:2018vaa}
G.~Bell, R.~Rahn and J.~Talbert, \emph{{Two-loop anomalous dimensions of
  generic dijet soft functions}},
  \href{https://doi.org/10.1016/j.nuclphysb.2018.09.026}{\emph{Nucl. Phys.}
  {\bfseries B936} (2018) 520}
  [\href{https://arxiv.org/abs/1805.12414}{{\ttfamily 1805.12414}}].

\bibitem{Bell:2018oqa}
G.~Bell, R.~Rahn and J.~Talbert, \emph{{Generic dijet soft functions at
  two-loop order: correlated emissions}},
  \href{https://doi.org/10.1007/JHEP07(2019)101}{\emph{JHEP} {\bfseries 07}
  (2019) 101} [\href{https://arxiv.org/abs/1812.08690}{{\ttfamily
  1812.08690}}].

\bibitem{Gatheral:1983cz}
J.~G.~M. Gatheral, \emph{{Exponentiation of Eikonal Cross-sections in
  Nonabelian Gauge Theories}},
  \href{https://doi.org/10.1016/0370-2693(83)90112-0}{\emph{Phys. Lett.}
  {\bfseries 133B} (1983) 90}.

\bibitem{Frenkel:1984pz}
J.~Frenkel and J.~C. Taylor, \emph{{Non-abelian eikonal exponentiation}},
  \href{https://doi.org/10.1016/0550-3213(84)90294-3}{\emph{Nucl. Phys.}
  {\bfseries B246} (1984) 231}.

\bibitem{Bell:2018jvf}
G.~Bell, R.~Rahn and J.~Talbert, \emph{{Automated Calculation of Dijet Soft
  Functions in the Presence of Jet Clustering Effects}},
  \href{https://doi.org/10.22323/1.290.0047}{\emph{PoS} {\bfseries RADCOR2017}
  (2018) 047} [\href{https://arxiv.org/abs/1801.04877}{{\ttfamily
  1801.04877}}].

\bibitem{Becher:2010tm}
T.~Becher and M.~Neubert, \emph{{{Drell-Yan} Production at Small $q_T$,
  Transverse Parton Distributions and the Collinear Anomaly}},
  \href{https://doi.org/10.1140/epjc/s10052-011-1665-7}{\emph{Eur. Phys. J.}
  {\bfseries C71} (2011) 1665}
  [\href{https://arxiv.org/abs/1007.4005}{{\ttfamily 1007.4005}}].

\bibitem{Becher:2011pf}
T.~Becher, G.~Bell and M.~Neubert, \emph{{Factorization and Resummation for Jet
  Broadening}},
  \href{https://doi.org/10.1016/j.physletb.2011.09.005}{\emph{Phys. Lett.}
  {\bfseries B704} (2011) 276}
  [\href{https://arxiv.org/abs/1104.4108}{{\ttfamily 1104.4108}}].

\bibitem{Chiu:2012ir}
J.-Y. Chiu, A.~Jain, D.~Neill and I.~Z. Rothstein, \emph{{A Formalism for the
  Systematic Treatment of Rapidity Logarithms in Quantum Field Theory}},
  \href{https://doi.org/10.1007/JHEP05(2012)084}{\emph{JHEP} {\bfseries 05}
  (2012) 084} [\href{https://arxiv.org/abs/1202.0814}{{\ttfamily 1202.0814}}].

\bibitem{Gangal:2014qda}
S.~Gangal, M.~Stahlhofen and F.~J. Tackmann, \emph{{Rapidity-Dependent Jet
  Vetoes}}, \href{https://doi.org/10.1103/PhysRevD.91.054023}{\emph{Phys. Rev.}
  {\bfseries D91} (2015) 054023}
  [\href{https://arxiv.org/abs/1412.4792}{{\ttfamily 1412.4792}}].

\bibitem{Becher:2011dz}
T.~Becher and G.~Bell, \emph{{Analytic Regularization in Soft-Collinear
  Effective Theory}},
  \href{https://doi.org/10.1016/j.physletb.2012.05.016}{\emph{Phys. Lett.}
  {\bfseries B713} (2012) 41}
  [\href{https://arxiv.org/abs/1112.3907}{{\ttfamily 1112.3907}}].

\bibitem{Hoang:2014wka}
A.~H. Hoang, D.~W. Kolodrubetz, V.~Mateu and I.~W. Stewart,
  \emph{{$C$-parameter distribution at N$^3$LL' including power corrections}},
  \href{https://doi.org/10.1103/PhysRevD.91.094017}{\emph{Phys. Rev.}
  {\bfseries D91} (2015) 094017}
  [\href{https://arxiv.org/abs/1411.6633}{{\ttfamily 1411.6633}}].

\bibitem{Becher:2012za}
T.~Becher, G.~Bell and S.~Marti, \emph{{NNLO soft function for electroweak
  boson production at large transverse momentum}},
  \href{https://doi.org/10.1007/JHEP04(2012)034}{\emph{JHEP} {\bfseries 04}
  (2012) 034} [\href{https://arxiv.org/abs/1201.5572}{{\ttfamily 1201.5572}}].

\bibitem{Becher:2012qc}
T.~Becher and G.~Bell, \emph{{NNLL Resummation for Jet Broadening}},
  \href{https://doi.org/10.1007/JHEP11(2012)126}{\emph{JHEP} {\bfseries 11}
  (2012) 126} [\href{https://arxiv.org/abs/1210.0580}{{\ttfamily 1210.0580}}].

\bibitem{Luebbert:2016itl}
T.~L{\"u}bbert, J.~Oredsson and M.~Stahlhofen, \emph{{Rapidity renormalized TMD
  soft and beam functions at two loops}},
  \href{https://doi.org/10.1007/JHEP03(2016)168}{\emph{JHEP} {\bfseries 03}
  (2016) 168} [\href{https://arxiv.org/abs/1602.01829}{{\ttfamily
  1602.01829}}].

\bibitem{Gangal:2016kuo}
S.~Gangal, J.~R. Gaunt, M.~Stahlhofen and F.~J. Tackmann, \emph{{Two-Loop Beam
  and Soft Functions for Rapidity-Dependent Jet Vetoes}},
  \href{https://doi.org/10.1007/JHEP02(2017)026}{\emph{JHEP} {\bfseries 02}
  (2017) 026} [\href{https://arxiv.org/abs/1608.01999}{{\ttfamily
  1608.01999}}].

\bibitem{Banfi:2012yh}
A.~Banfi, G.~P. Salam and G.~Zanderighi, \emph{{NLL+NNLO predictions for
  jet-veto efficiencies in Higgs-boson and Drell-Yan production}},
  \href{https://doi.org/10.1007/JHEP06(2012)159}{\emph{JHEP} {\bfseries 06}
  (2012) 159} [\href{https://arxiv.org/abs/1203.5773}{{\ttfamily 1203.5773}}].

\bibitem{Becher:2013xia}
T.~Becher, M.~Neubert and L.~Rothen, \emph{{Factorization and
  $N^{3}LL_{p}$+NNLO predictions for the Higgs cross section with a jet veto}},
  \href{https://doi.org/10.1007/JHEP10(2013)125}{\emph{JHEP} {\bfseries 10}
  (2013) 125} [\href{https://arxiv.org/abs/1307.0025}{{\ttfamily 1307.0025}}].

\bibitem{Stewart:2013faa}
I.~W. Stewart, F.~J. Tackmann, J.~R. Walsh and S.~Zuberi, \emph{{Jet $p_T$
  resummation in Higgs production at $NNLL'+NNLO$}},
  \href{https://doi.org/10.1103/PhysRevD.89.054001}{\emph{Phys. Rev.}
  {\bfseries D89} (2014) 054001}
  [\href{https://arxiv.org/abs/1307.1808}{{\ttfamily 1307.1808}}].

\bibitem{Frye:2016aiz}
C.~Frye, A.~J. Larkoski, M.~D. Schwartz and K.~Yan, \emph{{Factorization for
  groomed jet substructure beyond the next-to-leading logarithm}},
  \href{https://doi.org/10.1007/JHEP07(2016)064}{\emph{JHEP} {\bfseries 07}
  (2016) 064} [\href{https://arxiv.org/abs/1603.09338}{{\ttfamily
  1603.09338}}].

\bibitem{Borowka:2017idc}
S.~Borowka, G.~Heinrich, S.~Jahn, S.~P. Jones, M.~Kerner, J.~Schlenk et~al.,
  \emph{{pySecDec: a toolbox for the numerical evaluation of multi-scale
  integrals}}, \href{https://doi.org/10.1016/j.cpc.2017.09.015}{\emph{Comput.
  Phys. Commun.} {\bfseries 222} (2018) 313}
  [\href{https://arxiv.org/abs/1703.09692}{{\ttfamily 1703.09692}}].

\bibitem{Kardos:2020gty}
A.~Kardos, A.~J. Larkoski and Z.~Tr{\'o}cs{\'a}nyi, \emph{{Groomed jet mass at
  high precision}},  \href{https://arxiv.org/abs/2002.00942}{{\ttfamily
  2002.00942}}.

\bibitem{Kardos:2020ppl}
A.~Kardos, A.~J. Larkoski and Z.~Tr{\'o}cs{\'a}nyi, \emph{{Two- and Three-Loop
  Data for Groomed Jet Mass}},
  \href{https://arxiv.org/abs/2002.05730}{{\ttfamily 2002.05730}}.

\bibitem{Procura:2018zpn}
M.~Procura, W.~J. Waalewijn and L.~Zeune, \emph{{Joint resummation of two
  angularities at next-to-next-to-leading logarithmic order}},
  \href{https://doi.org/10.1007/JHEP10(2018)098}{\emph{JHEP} {\bfseries 10}
  (2018) 098} [\href{https://arxiv.org/abs/1806.10622}{{\ttfamily
  1806.10622}}].

\bibitem{Bell:2018gce}
G.~Bell, A.~Hornig, C.~Lee and J.~Talbert, \emph{{$e^+ e^-$ angularity
  distributions at NNLL$^\prime$ accuracy}},
  \href{https://doi.org/10.1007/JHEP01(2019)147}{\emph{JHEP} {\bfseries 01}
  (2019) 147} [\href{https://arxiv.org/abs/1808.07867}{{\ttfamily
  1808.07867}}].

\bibitem{Bell:2018mkk}
G.~Bell, B.~Dehnadi, T.~Mohrmann and R.~Rahn, \emph{{Automated Calculation of
  $N$-jet Soft Functions}},
  \href{https://doi.org/10.22323/1.303.0044}{\emph{PoS} {\bfseries LL2018}
  (2018) 044} [\href{https://arxiv.org/abs/1808.07427}{{\ttfamily
  1808.07427}}].

\end{thebibliography}\endgroup

\end{document}